\documentclass[journal,comsoc,10pt]{IEEEtran}
\usepackage[T1]{fontenc}% optional T1 font encoding
\usepackage{cite}

\ifCLASSINFOpdf
	\usepackage[pdftex]{graphicx}
\else
\fi

\usepackage{amsmath}
\usepackage{amssymb}
\usepackage{paralist}
\usepackage{enumitem}
\usepackage[cmintegrals]{newtxmath}
\usepackage{bm}
\usepackage{algorithmic}
\usepackage{algorithm}

\ifCLASSOPTIONcompsoc
\usepackage[caption=false,font=normalsize,labelfont=sf,textfont=sf]{subfig}
\else
\usepackage[caption=false,font=footnotesize]{subfig}
\fi

\usepackage{stfloats}
\usepackage{url}
\usepackage{booktabs,array}
\usepackage{tabularx}
\usepackage{epstopdf}
\usepackage{mathtools}
\usepackage{comment}
\usepackage{color}
\usepackage{colortbl}

\usepackage{pifont}
            
\definecolor{greycell}{gray}{.9}
\usepackage{mathrsfs}

\usepackage{xcolor, soul, colortbl}

\begin{document}
\title{Quality of Experience from Cache Hierarchies: Keep your low-bitrate close, and high-bitrate closer}
%\title{Cache Partitioning: A Road Towards Smooth Adaptive Streaming over ICN}
%\title{Cache Partitioning: Towards Smooth Adaptive Streaming over ICN}
%\title{Cache Partitioning: A Novel Paradigm for Smoother Adaptive Streaming over ICN}
% \title{Keep your low-bitrate close, and your high-bitrate closer: Partitioning Cache Hierarchies to Improve Quality of Experience}

\author{Wenjie~Li,~\IEEEmembership{Student Member,~IEEE,}
	Sharief~M.A.~Oteafy,~\IEEEmembership{Member,~IEEE,}\\
	Marwan~Fayed,~\IEEEmembership{Senior Member,~IEEE,}
	and~Hossam~S.~Hassanein,~\IEEEmembership{Fellow,~IEEE}
	\thanks{W. Li and H. S. Hassanein are with School of Computing, Queen's University, Kingston, ON, Canada. E-mail: \{liwenjie, hossam\}@cs.queensu.ca.}
	\thanks{S. Oteafy is with School of Computing, DePaul University, Chicago, IL, USA. E-mail: soteafy@depaul.edu.}
	\thanks{M. Fayed is with School of Computer Science, University of St Andrews, UK. E-mail: marwan.fayed@st-andrews.ac.uk.}}


\maketitle

\begin{abstract}
	Recent studies into streaming media delivery suggest that performance gains from cache hierarchies such as Information-Centric Networks (ICNs) may be negated by Dynamic Adaptive Streaming (DAS), the de facto method for retrieving multimedia content. The bitrate adaptation mechanisms that drive video streaming clash with caching hierarchies in ways that affect users' Quality of Experience (QoE). Cache performance also diminishes as consumers dynamically select content encoded at different bitrates. In this paper we use the evidence to draw a novel insight: in a cache hierarchy for adaptive streaming content, %low-bitrate content should be pushed into the network core, away from the edge, and prioritize capacity for high-bitrate content close to consumers. 
	bitrates should be prioritized over or alongside popularity and hit rates.
	We build on this insight to propose \emph{RippleCache} as a family of cache placement schemes 
	%that place content at routers along an entire forwarding path according to bitrates. Specifically, a \textit{RippleCache} scheme 
	that safeguard high-bitrate content at the edge and push low-bitrate content into the network core. Doing so reduces contention of cache resources, as well as congestion in the network. To validate \emph{RippleCache} claims we construct two separate implementations. We design \textit{RippleClassic} as a benchmark solution that optimizes content placement by maximizing a measure for cache hierarchies shown to have high correlation with QoE. In addition, our lighter-weight \textit{RippleFinder} is then re-designed with distributed execution for application in large-scale systems. \emph{RippleCache} performance gains are reinforced by evaluations in NS-3 against state-of-the-art baseline approaches, using standard measures of QoE as defined by the DASH Industry Forum.  Measurements show that \textit{RippleClassic} and \textit{RippleFinder} deliver content that suffers less oscillation and rebuffering, as well as the highest levels of video quality, indicating overall improvements to QoE.
\end{abstract}


\IEEEpeerreviewmaketitle

\section{Introduction}
\IEEEPARstart{T}{he} dominance of video traffic on the Internet makes it a high-value and high-priority candidate for in-network cache hierarchies such as Information-Centric Networks (ICNs)~\cite{westphal2016adaptive}. In the conventional IP-based Internet, streaming video traffic is known to defy the long-valued Internet tenets of stability, utilization, and fairness, in ways that are only beginning to be understood and addressed~\cite{Chen:2016,Mansy:2015,flach2016internet}. This suggests that video delivery services could be similarly problematic in cache hierarchies, 
%which is typically 
in particular when optimized to deliver non-adaptive and non-video traffic. ICNs being one such instantiation, it is therefore instructive to understand caching behaviour and design for adaptive media within the context of ICNs.

Dynamic Adaptive Streaming over HTTP (DASH) is the application-layer standard that is used to deliver multimedia content over the network~\cite{mpeg-dash}. A DASH implementation has three salient features. Content is first partitioned into equal duration segments. All segments are then encoded at multiple bitrates in order to accommodate a variety of network conditions. Finally, adaptation algorithms are used to retrieve the highest level of quality, subject to estimates of available network resources. These three attributes in combination have been central to maximizing consumer satisfaction, while minimizing costs of delivery for content providers. However, as streaming video traffic approaches 80\%~\cite{cisco2017}, application-layer solutions are facing issues of scale.

In-network caching of video segments with variable bitrates is touted as being one solution. The placement of video segments with variable bitrates in cache hierarchies, which is the subject of this paper, is known to be far from intuitive. Existing caching schemes (e.g., \cite{myjournal2017,ye2017quality}) fill this video-to-cache-placement gap by utilizing snapshots, or instantaneous inference, of adaptive video traffic in ICN. Despite some improvement, snapshots ignore the interplay between cache placement and consumer-side bitrate adaptation that can diminish cache performance~\cite{westphal2016adaptive}.

The challenges stem from the interaction between caches and bitrate adaptation that cause ``oscillation dynamics''~\cite{grandl2013interaction}. To exemplify a common scenario, consumers that retrieve low-bitrate segments from edge caches will perceive good performance. A consumer-side bit-rate adaptation protocol will thus invoke a request for higher-quality content that may be stored on a different (farther) cache in the network core. Data from the network core has to be delivered via a longer path than from the edge cache, and is more likely to face contention or congestion. Poor performance from the higher-quality video source will cause the streaming application to reduce its video quality preference. Oscillation dynamics are intrinsically linked to inaccurate estimates caused by ever-changing network conditions that occur with intermittent cache hits and misses. 
% In using snapshots, bitrate adaptation may recommend users to request for different bitrates, exceeding the expected behavior learned from the last snapshot of system. Cache placement that is derived from the last snapshot of system would become outdated immediately because of this shift on preferred bitrates.

%The issue of oscillation dynamics has been studied within the conventional Internet. For example, a cache-aware bitrate adaptation~\cite{lee2014caching} is proposed, where an independent thread of adaptation logic would be triggered when cache hits occur. However, ICN architecture significantly differs from conventional Internet where all content is hosted at known locations and consumer estimates of system performance are dominated by network effects. In an ICN where video segments with various bitrates may be stored in different caches, the consumer-side adaptation techniques have no means to distinguish poor performance in the network from poor performance at the cache. Thus, we argue that a ``good'' caching scheme that can stabilize the bandwidth fluctuation is the fundamental solution for ICN to relieve this oscillation.

Oscillation dynamics are not inherent to ICNs only, and have previously been studied in the context of Content Delivery Networks (CDN). For example, cache-aware bitrate adaptation~\cite{lee2014caching} triggers independent threads of adaptation logic when cache hits occur.  However, caching in CDNs differs significantly from cache hierarchies in ICNs. CDNs host all video content, and at fixed locations, so consumer estimates of system performance are dominated by network effects. In contrast, cache hierarchies in ICNs make it possible for video segments to appear at any cache router. As a result, consumer-side adaptation techniques have no means to distinguish between poor performance from network conditions and poor performance from cache conditions. This suggests that a ``good'' caching scheme may stabilize bandwidth fluctuations to reduce oscillation, and thereby improve consumer Quality of Experience (QoE). 

In this paper, we posit that one such family of caching schemes emerges when encoding bitrates are prioritized over - or alongside - conventional metrics associated with hit rates and popularity. In particular, we hypothesize that the QoE for high-quality content requests suffers dis-proportionally from resource sharing, relative to low-quality content. One implication would be that the highest bitrate content should be placed where there is least congestion. Our investigations into adaptation-based caching dynamics show that bitrate oscillation patterns emerge with hop distance~\cite{ourletter,li2018bitrate}. The pattern that emerges suggests that high-bitrate content is most stable when retrieved from edge caches. From a caching perspective this may be counter-intuitive: it entails copies of the largest segments at multiple edge caches, rather than a single copy at upstream caches that sit on intersecting paths.

This insight leads to, and is validated by, the main contribution of this paper in \textit{RippleCache}. We present \textit{RippleCache} as a cache guiding principal that safeguards capacity at the edge routers for high-bitrate content, thereby pushing lower bitrate content along the forwarding path towards the network core.
% content first use initial evaluations to understand the influence of adaptive streaming on cache placement. This insight then motivates us to safeguard cache capacity, which forms cache partitions across each forwarding path according to available bitrates.  As a result, video content at a selected bitrate can be retrieved within an appropriate hop distance that stabilizes bandwidth fluctuation.
This has the effect of \textit{partitioning} cache capacity along a forwarding path, but raises questions with respect to partition boundaries and caches that sit on intersecting paths. In order to validate the main contribution, we construct two independent \emph{RippleCache}-guided systems:
% Our contributions in this work are listed as follows:
\begin{enumerate}
% 	%\item We carry out exploratory experiments on existing caching schemes to observe the cache distribution under adaptive streaming, and the observed interplay between caches along a forwarding path motivates the idea of a single aggregate of caches into a \textit{cache path}.
% 	\item We introduce a notion of \textit{\textbf{RippleCache}}, as a guiding principle for adaptation-aware cache partitioning. \textit{RippleCache} views entire cache capacity across each forwarding path into a \textit{cache path}, and keeps high-bitrate content on the edge of cache path close to consumers, while pushing low-bitrate content into the network core.
	
\item  \textit{\textbf{RippleClassic}} serves as a benchmark cache partitioning paradigm. Partitions are created by solving an optimization problem formulated as binary integer programming. The objective of \textit{RippleClassic} maximizes a metric designed specifically to measure cache hierarchy performance for adaptive streaming, that has been shown to have high correlation with consumers QoE~\cite{ourletter}. The solutions that emerge place content in such a way that a \emph{RippleCache} emerges.
	
\item \textit{\textbf{RippleFinder}} is a distributed caching scheme that is built on our prior work~\cite{li2018bitrate} and executes in polynomial-time complexity. Execution begins at edge routers, from where cache partitions are created along the forwarding path to each video producer. Placement decisions prioritize utility, an indicator of the resource cost of a video segment (by size) and weighted by popularity.
\end{enumerate}

Performance evaluations compare both \emph{RippleCache} designs with ProbCache~\cite{psaras2012probabilistic} as a baseline for probabilistic caching, as well as CE2~\cite{zhang2015survey} as a baseline that commonly appears in literature. Measures are selected and defined in accordance with DASH Industry Forum recommendations~\cite{dashmetrics}. \emph{RippleCache} constructions consistently reduce oscillation and re-buffering, while meeting or exceeding the highest levels of competing video quality. The consistent performance, across varying levels of capacity and popularity-skew, lend weight to the argument that  high-bitrate content should be kept close to consumers, and lower quality content pushed further away.

%We instantiate our partitioning framework in, and evaluate its performance using, NS-3 based ICN simulator ndnSIM~\cite{alexander2012ndnsim}. FESTIVE~\cite{jiang2012improving} was also implemented to provide bitrate adaptation. The performance of \textit{RippleClassic} and \textit{RippleFinder} is evaluated against Least Recently Used (LRU)- and Least Frequently Used (LFU)-based \emph{Cache Everything Everywhere (CE2)}~\cite{zhang2015survey} and \emph{ProbCache}~\cite{psaras2012probabilistic}. Observations suggest both \textit{RippleClassic} and \textit{RippleFinder} ensure high video quality comparing with popularity-based schemes, as well as significant reductions in video playback freezing and bitrate oscillation.

The remainder of this paper is organized as follows. In Section~\ref{Sec:RelatedWork} we present related work, focusing on recent contributions to bitrate adaptation control and video caching in ICNs. Section~\ref{Sec:RippleCache} pinpoints the challenges of adaptation-agnostic caching schemes on adaptive video streaming, and presents the \textit{RippleCache} principle. To assess the potential gain of \textit{RippleCache}, We formulate a benchmark solution \textit{RippleClassic} in Section~\ref{Sec:Optimization}, followed by a light-weight and practical embodiment \textit{RippleFinder} in Section~\ref{Sec:RippleFinder}. Section~\ref{Sec:Performance} presents our experiment setup and performance evaluation. We conclude in Section~\ref{Sec:Conclusion} and present our final remarks.
\section{Related Work}\label{Sec:RelatedWork}
Ubiquitous caching~\cite{zhang2015survey} is a fundamental feature of ICN, and could effectively reduce redundant traffic generated by duplicate requests. Due to the decoupling of content and location in ICN naming mechanisms, information is not bound to a certain host, and can be retrieved from anywhere in the network. In-network caching schemes in ICN have been heavily investigated~\cite{ioannou2016survey}. A consensus is reached where caching performance can be enhanced by catering to content popularity~\cite{li2012popularity,cho2012wave}. For example, request statistics may be processed to make caching decisions that reduce the hop distance between consumer and content~\cite{li2012popularity}. The request frequency has also been utilized to annotate segments of popular content and resize caching windows~\cite{cho2012wave}.

In the domain of adaptive video streaming, users' QoE can be improved by both client-side and server-side control~\cite{kua2017survey}. Rate adaptation on the client-side can be \textit{Throughput}-based~\cite{li2014probe,jiang2012improving}, or \textit{Buffer}-based~\cite{huang2015buffer,tian2012towards}. \textit{Throughput}-based adaptation makes the best possible estimates on bandwidth by referring to received video throughput from previous segments, and adjusts bitrate selections to match bandwidth estimates. \textit{Buffer}-based adaptation, is instead guided by indirect means of resource estimation such as buffer occupancy. %Our proposed cache partitioning schemes cooperate with \textit{Throughput}-based adaptation control, as we manipulate video throughput in order to deliver smooth playback.}

%In addition to studies which support ubiquitous in-network caching, recent proposals suggest that edge-caching, alone, may offer performance improvement that is equivalent to ubiquitous in-network caching~\cite{fayazbakhsh2013less}. However, this observation is constrained to LRU cache replacement of non-video content. Our observations in Section~\ref{Sec:Performance} suggest that allocating more cache capacity on the edge under-performs against equal-size cache allocation when delivering video content of variable bitrates.

The relationship between in-network caching and bitrate adaptation has also attracted attention. For example, Jia et al.~\cite{jia2016modeling} designed a control layer for optimal Interest forwarding and adaptive video caching based on the virtual queue of each bit rate. Kreuzberge et al.~\cite{kreuzberger2015modelling} develop a cache-aware traffic-shaping policy in response to the unfair bandwidth sharing generated by rate-adaptive video streams. Liu et.al.~\cite{liu2013dynamic} studied caching behavior over ICN and demonstrated that clients could be served with bit rates even higher than their actual bandwidth, emphasizing the potential of ICN caching and the importance of designing caching schemes for adaptive video content. Other studies, such as our previous work on revealing the interplay between caching and bitrate adaptation~\cite{li2017performance}, motivates the need of bitrate-adaptation-aware caching. The focus on caching specifically for adaptive video streaming is comparatively recent. Examples include building cache models that accommodate multiple bitrates of the same content~\cite{myjournal2017,jin2015towards,araldo2016representation}. However, these works either drive caching mechanisms using the steady states that emerge from modelling bitrate adaptation as a Markovian process~\cite{myjournal2017}, or assume random behaviour from bitrate adaptation generated by Gaussian model~\cite{jin2015towards}. While insightful, these studies are built on assumptions that overlook the real-world variations of client-side bitrate adaptations.

In this work, we specifically address the interaction between ubiquitous caching and bitrate adaptation. To the best of our knowledge, our proposed \textit{RippleCache} principle, and its embodiments, are the first attempts to address the interplay between bitrate adaptation and cache placement to improve users' QoE.
\section{Adaptive Streaming with Cache Partitioning}\label{Sec:RippleCache}
Our initial studies on the interplay between consumer-side adaptation and in-network cache placement provide us an opportunity to observe adaptation-level dynamics that vary by hop distance. These observations then demonstrate the need for safe-guarding cache capacity for a particular bitrate to facilitate fine-grained cache placement, which lead to our design of the \textit{RippleCache} principle.

\subsection{Adaptation Dynamics Characterized by Cache Placement}
To study the impact of consumer-side bit-rate adaptation on cache placement, we carried out extensive experiments to elicit the intrinsic challenge of bitrate oscillation and high bit-rate placement. The following characterizations are drawn from evaluations of the benchmark Cache Everything Everywhere (CE2) with Least Frequently Used (LFU)~\cite{zhang2015survey}. The evaluation setup is described in Section~\ref{Sec:Performance}.
\begin{figure}[!t]
	\centering
	\includegraphics[width=0.95\columnwidth]{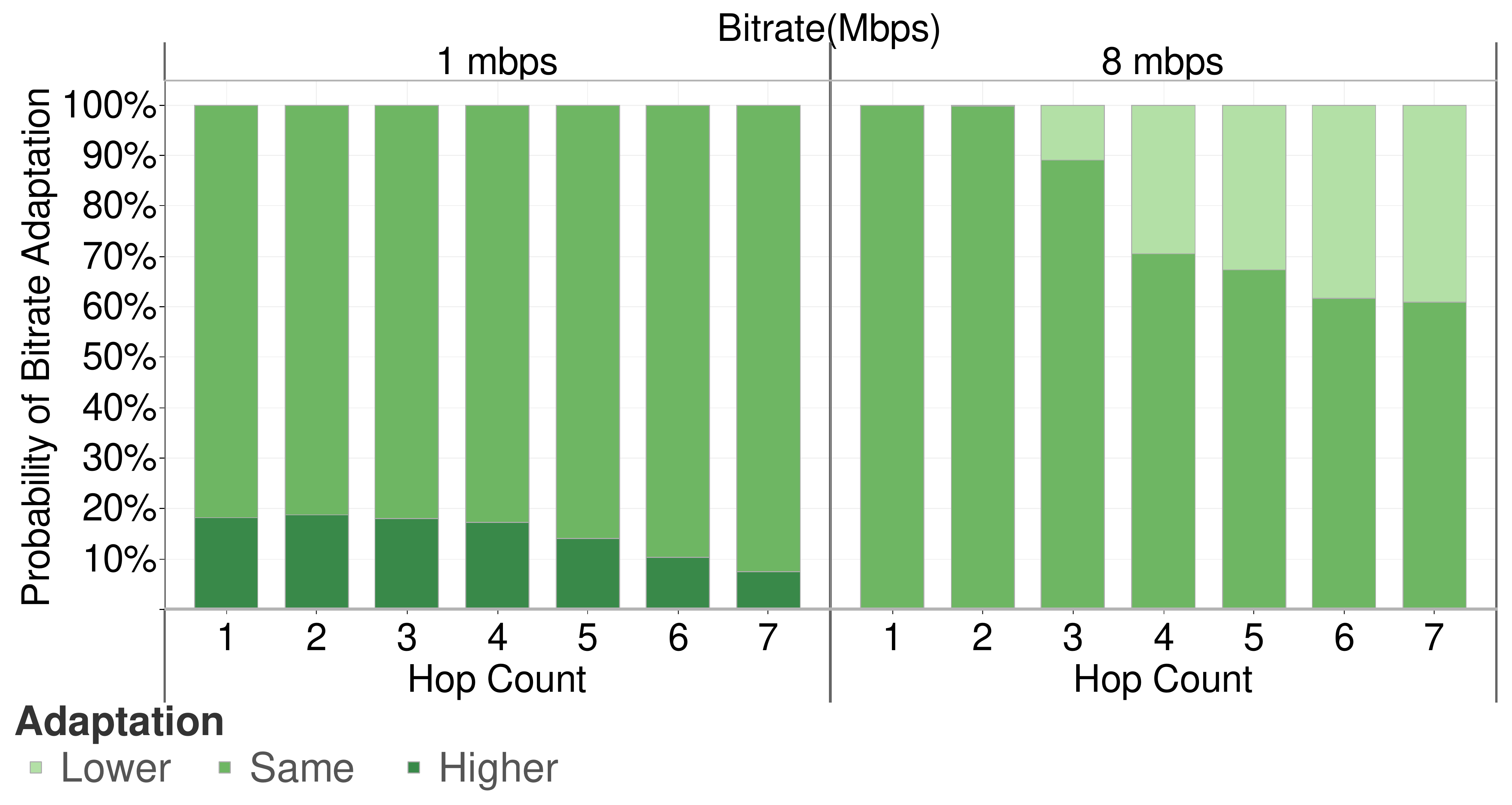}
	\caption{Bitrate adaptations given cache distance: dark regions indicate switches to the higher bitrate; lighter regions indicate switches to the lower bitrate.}
	\label{Fig:CacheSwitch}
\end{figure}

The salient results are summarized by Figure~\ref{Fig:CacheSwitch}, depicting the likelihood of incurring a bitrate adaptation as a function of hop distance between the video consumer and the cache. %This is an important measure because bitrate oscillations is known to be a strong source of dissatisfaction, \textit{irrespective and independent of selected bitrate}~\cite{Dobrian2011}.
Each vertical bar is shaded according to the the direction of the adaptation: dark regions indicate switches to a higher bitrate; lighter regions indicate switches to lower bitrates; medium shade indicates no bitrate adaptation (same decision). We note that bitrate adaptations may be triggered in response to changes in either or both of network and caching conditions. Thus, the proportion of medium shade is an indication of stable or steady state between video requests with the network and caches that satisfy those requests. In order to reduce bitrate oscillation, this proportion of medium shade is expected to be as more as possible.

Bitrate adaptations occur most frequently relative to cache distances when users consume the lowest (1 mbps) and highest (8 mbps) bitrates under our experimental settings. As depicted in Figure~\ref{Fig:CacheSwitch}, the left-most bars show bitrate adaptations after successful requests for video content at 1 mbps. From among requests for low bitrates satisfied within the first four hops, measurements indicate no significant difference in the likelihood of a bitrate increase. This suggests a degree of insensitivity to the location of low-bitrate content, with no obvious advantage to caching low-bitrate content closer to consumers at the edge. Instead, caching low-bitrate content in the core network provides an increasing adaptation stability, as proportion of medium shade increases in the last three hops.

In contrast, the rightmost bars in Figure~\ref{Fig:CacheSwitch} show an opposing trend. Consumers that request high-bitrate content are increasingly likely to switch to lower quality as hop distance increases. Service degradation becomes increasingly unavoidable with hop distance for high-bitrate content. This happens because higher bitrate content consumes a disproportionately greater share of cache and network resources.

The combination of these two sets of observations suggest that lower-bitrate content should be moved into the core to make room for higher-bitrate content at the edges, which demonstrates the need for safe-guarding cache capacity for a particular bitrate. These observations then motivate our design of an adaptation-aware cache partitioning to reduce bitrate oscillation and improve users' QoE.

\subsection{Ripple-like Cache Partitioning}
\begin{figure}[!t]
	\centering
	\includegraphics[width=0.95\columnwidth]{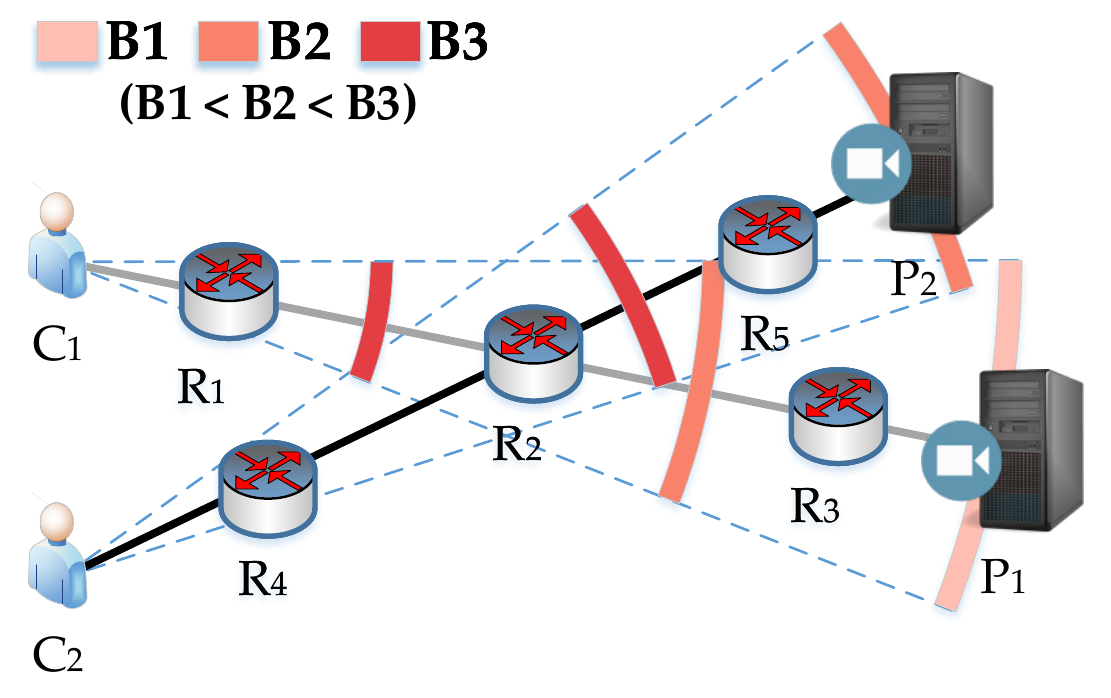}
	\caption{Cache partitioning by encoding bitrates along each forwarding path.}
	\label{Fig:CR}
\end{figure}
Our early experiments underscore the need for cache partitioning. However, rather than conventional partitioning on individual cache, we propose \textbf{RippleCache} partitioning principle that works upon each \textit{cache path}. A \textit{cache path} is a concatenation of caches that sit on a forwarding path from consumers to a video producer. We say that a \textit{RippleCache} principle safeguards content along the cache path by prioritizing bitrates in a monotonically decreasing fashion from edge routers.
% as follows,
% \begin{description}
% \item [RippleCache.] \textit{Caches along each \textit{cache path} are safeguarded for particular bitrates in a monotonically decreasing fashion from edge routers.}
% \end{description}

The bitrate assignments in \textit{RippleCache} effectively partition caches into concentric regions that we refer to as \textbf{Ripple}s. Ripple behaviours derive from its namesake: As much as ripples in liquid ebb and flow, partitions must be dynamic or re-definable in response to changing interest patterns.

A visual representation of a \textit{RippleCache} is given in Figure~\ref{Fig:CR}. It shows two independent forwarding paths from consumers $C_1$ and $C_2$ to their respective video producers $P_1$ and $P_2$. Caches, as guided by \textit{RippleCache}, are assigned one of three available bitrates $B_3 > B_2 > B_1$, in decreasing bitrate from the consumers. The coloured arcs in Figure~\ref{Fig:CR} mark the partition boundaries that delineate \textit{Ripple}s. We note that \textit{Ripple}s may contain 0 or many routers. For example, the path from $C_2$ to $P_2$ assigns bitrate $B_3$ to the two routers closest to $C_2$; the same path omits the lowest bitrate from its partitions, leaving the video producer to satisfy lowest bitrate requests.

%The forwarding paths in Figure~\ref{Fig:CR} also intersect at $R_2$. This creates a conflict at $R_2$ where cache capacity must be shared. Equal share is one na\"{i}ve solution, given that $R_2$ is equidistant from the edge routers on the intersecting paths. However, the two intersecting partitions that contain $R_2$ are also allocated different bitrates. Even if controls exist for interest and popularity of content along each path, equal share of capacity ignores that higher bitrate content consumes dis-proportionally greater resources.
Same as ripples in liquid must coincide when they meet, caches that sit on multiple forwarding paths must share their capacity to resolve potential conflicts on cache partitions. For example, the forwarding paths in Figure~\ref{Fig:CR} intersect at $R_2$, where cache space is reversed for different bitrates from these two paths: the same router $R_2$ is requested to cache both $B_3$ and $B_2$. As a result, a spontaneous solution is dividing the cache space at $R_2$ to ensure a fair share among these two \textit{cache path}s, such that video content with $B_2$ and $B_3$ can coincide.

Our proposed \textit{RippleCache} provides a manifestation of the `ideal' cache partitioning. However, it is still a guiding principle and must be realized by a caching scheme in practice. A \textit{RippleCache} implementation must
\begin{inparaenum}
	\item identify appropriate caching decision criteria so that placements may form partitions; and
	\item implement a negotiation mechanism to ensure fair share allocations of cache capacity at nodes on intersecting paths. 
\end{inparaenum} 
The following sections describe our implementations in \emph{RippleClassic} as a benchmark and \emph{RippleFinder} as a scalable and distributed heuristic.

\section{\textit{RippleClassic} Benchmark Optimization}\label{Sec:Optimization}
% How RippleClassic works // move something from RippleFinder: (Execution, central controller<--hahaha)
% How these two issues are resolved.....cache decision criteria: cache reward, conflicting partitioning: optimization.
Guided by the \textit{RippleCache} principle described in the previous section, we hereby present the \textit{RippleClassic} cache placement scheme. \textit{RippleClassic} is an optimization formulated as a binary integer programming (BIP) problem. Its solutions are cache placements for adaptive video content under diverse network conditions and preferences. These placements serve as the benchmarks, against which we design and compare in later sections.

\subsection{Cache Placement Problem Formulation}\label{Sec:sub:formulation}
We model an ICN as a connected graph $G = (\mathbb{V}, \mathbb{E})$, where nodes in $\mathbb{V}$ are composed of video producers $\mathbb{P}$, edge routers $\mathbb{D}$ and intermediate routers. Each user is served exclusively by one edge router. Every node $v \in \mathbb{V}$ is equipped with content storage capacity $C_v$ dedicated to adaptive video caching. In this formulation single-path forwarding is assumed, where routers satisfy video requests by selecting the path with least delay to deliver content. Our formulation then optimizes the contributions of in-network caching along each forwarding path individually.

The number of video files in the system is represented by $F$. Our model reflects that content for adaptive streaming is fragmented into equal \emph{duration} segments, i.e., segments encoded at variable bitrate will be variable sizes. For ease of presentation, video files are fragmented into the the same number of fragments $K$. The number of bitrate encodings is $B$. Hence, video segments are identified by a \emph{(file, segment, bitrate)} triple, $(f, k, b)$, where $1\leq f\leq F, 1\leq k\leq K$, and $1\leq b\leq B$. Each video segment has size $S(f,k,b)$; we use $S(b)$ to simplify notation since equal duration video segments vary in size with bitrate encodings.

Let $x_v$ denote the cache placement decision, where $v \in \mathbb{V}$ and $x_v(f, k, b) \in\{0,1\}$. Thus a decision of $x_v(f, k, b) = 1$ indicates that video segment $(f, k, b)$ is cached at node $v$. We define $[d,p]$ as the sequence of routers on the forwarding path from edge router $d \in \mathbb{D}$ to the video producer $p \in \mathbb{P}$. The length of this forwarding path $[d,p]$ is $L$ and the index of $[d,p]$ starts from 1. Thus $x_{[d,p]}^i$ represents the cache decision variable on the $i^{th}$ router of $[d,p]$ (where $f$, $k$ and $b$ are implicit). Each $x_{[d,p]}^i$ also becomes an alias within $x_v$ for example, $x_{[d,p]}^1$ is an alias of $x_d$, which provides a view of the edge router on the forwarding path to $p$.

We further define binary variable $\delta_{[d,p]}^i$ as the \textit{caching status indicator}, which reflects an `aggregated' cache placement decision from $d$ to $i^{th}$ router of $[d,p]$. $\delta_{[d,p]}^i = 1$ only if any cache placement decision variable $x_{[d,p]}^j = 1$ for $1\leq j \leq i$. In other words, $\delta_{[d,p]}^i = 1$ if content is already cached on a downstream router. Finally, the number of requests on video segment $(f,k,b)$ received by edge router $d$ is denoted as $\theta_d(f,k,b)$, with $\theta_d$ substituted for simplicity. Notation is additionally summarized in Table~\ref{Table:Notation}.
\begin{table}[!tbp]
	\caption{Summary of notations used in the formulation}\label{Table:Notation}
	\begin{center}
		\begin{tabular}{ll}
			Notation & Meaning\\
			\toprule
			$\mathbb{V}$ & Set of ICN nodes\\
			$\mathbb{E}$ & Set of links\\
			$\mathbb{D}$ & Set of edge routers\\
			$\mathbb{P}$ & Set of video producers\\
			$S$ & Sizes of video segments\\
			$C$ & Cache capacity of ICN router\\
			$B$ & Number of supported bitrates\\
			$F$ & Number of adaptive video files\\
			$K$ & Number of video Segments in any file\\
			$x$ & Cache placement decision\\
			$[d,p]$ & ICN routers on forwarding path from edge $d$ to producer $p$\\
			$\delta$ & Caching status indicator (0 or 1)\\
			$\theta$ & Number of video requests received by edge router\\
			$\gamma$ & Cache reward value\\
			$RB_{[d,p]}^i$ & Ripple Bitrate on $i^{th}$ router along $[d,p]$\\
			\bottomrule
		\end{tabular}
	\end{center}
\end{table}

Our formulation caters to diverse caching preferences by maximizing the sum of cache reward values. The cache reward of each request is denoted by $\gamma(RB_{[d,p]}^i,b)$, generated by a reward function that is described subsequently in Section~\ref{Sec:Protocol}. The optimization is formulated as a BIP problem, as outlined below. Given the known complexity of BIP, solving this problem is NP-Complete.
\begin{alignat}{2}
\max\enskip & \sum_{d\in\mathbb{D}}\sum_{p\in\mathbb{P}}\sum_{i=1}^L\sum_{f = 1}^F\sum_{k = 1}^K\sum_{b = 1}^B\gamma(RB_{[d,p]}^i,b)\theta_d[\delta_{[d,p]}^i-\delta_{[d,p]}^{i-1}]\nonumber\\
\mbox{s.t.}\quad
	&x_v(f,k,b) \in \{0,1\},\quad \forall v \in \mathbb{V}\label{Eq: XBinary}\\
	&\delta_{[d,p]}^i \in \{0,1\},\quad\forall d \in \mathbb{D}, \forall p \in \mathbb{P}, 1\leq i \leq L \label{Eq:GammaBinary}\\
	&\sum_{f\in F}\sum_{k\in K}\sum_{b\in B} S(b)*x_v(f,k,b) \leq C_v,\hspace{1pt}\forall v \in \mathbb{V} - \mathbb{P}\label{Eq: Capacity}\\
	&\delta_{[d,p]}^i \geq \delta_{[d,p]}^{i-1}, \label{Eq: Relation1}\\
	&\delta_{[d,p]}^i \geq  x_{[d,p]}^i(f,k,b), \label{Eq: Relation2}\\
	&\delta_{[d,p]}^i \leq  \delta_{[d,p]}^{i-1} +  x_{[d,p]}^i(f,k,b),\label{Eq: Relation3}\\
	&\delta_{[d,p]}^0 = 0,\label{Eq: SpecialCondition2}\\
	&x_p(f,k,b) = 1,\quad\forall p \in \mathbb{P}\label{Eq: SpecialCondition1}\\
	\begin{split}
		\delta_{[d,p]}^{L-1}(f',k',b)-\delta_{[d,p]}^i(f',k',b)&\\
		\leq \mathbf{M}- \mathbf{M} * \delta_{[d,p]}^i(f,k,b),& \\
	\end{split}\label{Eq: Popularity}
\end{alignat}

\subsubsection{Objective}
%Our proposed binary integer programming formulation 
The objective function maximizes the system-wide cache reward. A higher cache reward value correspond to a better cache placement. The optimization traverses all forwarding paths starting from each edge router, and accumulates cache reward values on nodes where cache hits occur. Cache rewards are generated once per request where the cache hit occurs. The objective expression thus utilizes the difference between cache indicators $\delta$ to avoid infeasible reward values: in cases where a segment is cached multiple times along the forwarding path, $\delta_{[d,p]}^i$ and $\delta_{[d,p]}^{i-1}$ would be both equal to 1. Their difference $\delta_{[d,p]}^i-\delta_{[d,p]}^{i-1}$, being 0, ensures the correctness of reward calculation. Only where the segment first appears along the path can rewards be accumulated, i.e.\@ where $\delta_{[d,p]}^i-\delta_{[d,p]}^{i-1}$ is non-zero.
%Since a video request successfully reaches router $L_{n, \mathbb{P}}(i)$, it inherently means cache misses on all nodes from $L_{n, \mathbb{P}}(0)$ to $L_{n, \mathbb{P}}(i-1)$, such that $\delta(n,i-1)$ is set to 0 and $\delta(n,i)-\delta(n,i-1)$ is 1. Similarly, as to upstream router $j$ (where $j > i$) on path $L_{n, \mathbb{P}}$, $\delta(n,j)$ and $\delta(n,j-1)$ are both equal to 1 because $x_{L_{n, \mathbb{P}(i)}}(f,k,b)=1$ for their common downstream node $L_{n, \mathbb{P}}(i)$, which ensures $\delta(n,i)-\delta(n,i-1) = 0$ and avoids accumulating infeasible cache reward.

\subsubsection{Constraints}
Binary variables are defined in Constraints (\ref{Eq: XBinary}) and (\ref{Eq:GammaBinary}). The remaining constraints relate to the \textit{Cache Capacity}, \textit{Caching Status Indicator}, and \textit{Popularity}, as follows.

\begin{inparaenum}[\textbullet]
	\item The \textit{Cache Capacity} defined in Constraint (\ref{Eq: Capacity}) ensures that the total size of cached video content is bound by available cache capacity over all cache routers except video producer.
	
    %In our problem formulation, 
	%The optimal result of caching decision variable $x$ guides cache placement, while our optimization objective utilizes a summarized caching status variable $\delta$ instead of $x$ to avoid infeasible cache rewards.
	\item The relationship between \textit{Caching Status Indicator} $\delta$ and cache placement decisions $x$ is defined by Constraints (\ref{Eq: Relation1})-(\ref{Eq: SpecialCondition2}). $\delta$ is an aggregation of cache placement decisions $x$. Constraints (\ref{Eq: Relation1}) and (\ref{Eq: Relation2}) give the lower bound of $\delta_{[d,p]}^i $, ensuring that its value should be greater than or equal to both its last hop indicator $\delta_{[d,p]}^{i-1}$ and the cache placement decision of the current router $x_{[d,p]}^i$. Constraint (\ref{Eq: Relation3}) gives the upper bound. When both $\delta_{[d,p]}^{i-1}$ and $x_{[d,p]}^i $ are 0, Constraint (\ref{Eq: Relation3}) will enforce $\delta_{[d,p]}^i$ to be assigned 0 since video content is not cached yet along $[d,p]$. Constraints (\ref{Eq: SpecialCondition2}) and (\ref{Eq: SpecialCondition1}) cope with the two special cases that are the consumer and the producer, respectively. As the index of $[d,p]$ starts from $i = 1$, we assign $\delta_{[d,p]}^0 = 0$. Conversely, the caching decision on video producer $x_p$ is equal to 1 for any content, since unavailable content on in-network caches can always be found at the producer.
	
	\item \textit{Popularity} contributes via Constraint (\ref{Eq: Popularity}). 
	The catering to popularity is known to improve the performance of caching schemes~\cite{cho2012wave, myjournal2017,zhang2015survey}.
	%Previous work ~\cite{cho2012wave, myjournal2017,zhang2015survey}) show better performance of caching schemes when catering to popular content. 
	Constraint (\ref{Eq: Popularity}) ensures that, whenever there is cache space, popular video content is selected for caching with a higher priority (close to consumers). We utilize the `big-$\mathbf{M}$' approach~\cite{griva2009linear} to ensure caching order,
	%in Constraint (\ref{Eq: Popularity}), 
	where $\mathbf{M}$ is any large positive constant number. 
	%Specifically, if video content $(f',k',b)$ is more popular than content $(f,k,b)$ and content $(f,k,b)$ is cached on $i^{th}$ router of $[d,p]$, then $(f',k',b)$ cannot be cached on any router $j$ where $ i < j \leq L$.
	
	The popularity Constraint (\ref{Eq: Popularity}) benefits from additional remarks. A ranking table is assumed to exist for each forwarding path; in our own implementation (described in Section~\ref{Sec:Performance}) ranking tables are held and maintained at each edge routers $d$. Entries in the table are first categorized into bitrates, and then sorted by popularity for each category. We denote $(f',k',b)$ and $(f,k,b)$ as any two consecutive items in this table, where segment $(f',k',b)$ is more popular than content $(f,k,b)$. Constraint~(\ref{Eq: Popularity}) guarantees that a less popular $(f,k,b)$ cannot be cached closer to consumers than $(f',k',b)$ on the forwarding path $[d,p]$. The result of left hand side is 1 if $(f',k',b)$ is cached on any upstream router from $i$ to penultimate node of the path $[d,p]$. To ensure~Constraint (\ref{Eq: Popularity}) is not violated, $\delta_{[d,p]}^i(f,k,b)$ in the right hand side must then be assigned 0, which represents that $(f,k,b)$ is never cached on a $j^{th}$ downstream router closer to consumers ($1 \leq j \leq i$).
\end{inparaenum}

\subsection{Cache Reward Function}\label{Sec:Protocol}
% The defined objective maximizes the aggregate of some measure cache reward, $\mu$, that is generated whenever a request is satisfied by a cache. As an indicative measure of quality of experience, RippleClassic adopts a reward function that we first introduced in~\cite{ourletter}. It is instructive to summarize its design here, so that its use and further refinement in the context of RippleClassic may be described.

\textit{RippleClassic} decides content placement among caches, without explicit knowledge of the interplay between caches and consumers. This information is encoded in and modelled by the reward function $\gamma$. The design of $\gamma$ relies on the following intuition: A cache hit is valuable only if the transfer of video content from that cache to the consumer can be reliably sustained for the requested bitrate. 

This intuition is demonstrated by an example in Table~\ref{Table:RippleBitrate}. 
Each cell in the table is the average transfer delay for a 4-second video segment delivered to consumer $C$ from any cache on the forwarding path. The greyed cells delineate the routers from which content can be reliably retrieved within 4-second time for a given encoding. The duration of a video segment (4 seconds) is the deadline for video delivery: meeting this deadline means the requested bitrate is reliable, while missing this deadline may ultimately cause video playback freezing. For example, video requests for $B_2$ are only sustainable when satisfied on $R_1$ or $R_2$; video content retrieved from $R_3$ or $R_4$ will arrive 2.5s or 7s late on average.

Transfer delay also gives an indication on the \textit{value} of a cache hit to the consumer. Referring again to Table~\ref{Table:RippleBitrate}, the delineation by greyed cells also corresponds with consumer adaptations. Consider a consumer that selects content encoded into 4-second segments at bitrate $B_2$. Table~\ref{Table:RippleBitrate} says that consumers would maintain or even increase their selected bitrate when content retrieved from $R_2$ or $R_1$. Conversely, that same content retrieved from $R_3$ or $R_4$ will cause the consumer to avoid playback freezing by reducing its selected bitrate. This type of oscillation is the behaviour observed in Figure~\ref{Fig:CacheSwitch}.

\begin{table}[!tbp]
	\centering
	\caption{An Example of Average Cumulative Delay of 4-second segments by Hop distance}
	\label{Table:RippleBitrate}
	%\resizebox{.48\textwidth}{!}{
	\begin{tabular}{l|cccc} \toprule
		&$(C..R_1)$&$(C..R_2)$&$(C..R_3)$&$(C..R_4)$\\\hline
		$B_3$& \multicolumn{1}{>{\columncolor{greycell}}c}{3s} & 6.5s & 10.5s & 16.5s\\
		$B_2$& \multicolumn{1}{>{\columncolor{greycell}}c}{1s} & \multicolumn{1}{>{\columncolor{greycell}}c}{3.5s} & 6.5s &11s\\ 
		$B_1$& \multicolumn{1}{>{\columncolor{greycell}}c}{0.5s} & \multicolumn{1}{>{\columncolor{greycell}}c}{1s} & \multicolumn{1}{>{\columncolor{greycell}}c}{2s}&\multicolumn{1}{>{\columncolor{greycell}}c}{3s}\\
		\bottomrule\end{tabular}%
	%}
\end{table}

The consumer-side adaptation and its interaction with in-network caches are then captured by reward function $\gamma$ that we first introduced in~\cite{ourletter}, where the numerical reward values were shown to have a high correlation with traditional consumer-side measures of QoE. The function takes two input parameters: 
\begin{inparaenum}[(i)]
  \item the consumer's requested bitrate $b$ and, crucially,
  \item the router's assigned Ripple Bitrate, ($RB_{[d,p]}^i$).
\end{inparaenum} 
Given the $i^{th}$ router on the forwarding path from edge node $d$ to producer $p$, \textbf{Ripple Bitrate} ($RB_{[d,p]}^i$) denotes the \textit{highest} sustainable bitrate that can be delivered to consumers (i.e., the top greyed cell of each column in Table~\ref{Table:RippleBitrate}). We further use $RB^i$ to denote $RB_{[d,p]}^i$, where forwarding path $[d,p]$ is implicit.

The reward function $\gamma(RB^i,b)$ is defined as,
\begin{equation}
\gamma(RB^i,b)= \left\{
\begin{split}
&\mu(b)\text{,} \quad\textbf{if} \ \ b= RB^i\\
&\mu(b^\uparrow) * \beta(b) + \mu(b) * (1-\beta(b))\text{,} \hspace{3pt}\textbf{if} \ \ b < RB^i\\
&\mu(RB^i)\text{,} \quad\textbf{if} \ \ (b> RB^i) \land (RB^i \geq b_1)\\
&\mu(b_1), \quad\textbf{otherwise.}\\
\end{split}
\right.
\end{equation}

We note that storage and transmission requirements for the encodings of any single video segment are non-uniform. In order to ensure that similar bias is reflected in the reward, $\mu$ is proportional to the base segment size. For the base bitrate at rank 1, $\mu(b_1) = 1$. Any other bitrate $b$ is calculated as $\mu(b) = S_b/S_{b_1}$, where $S_{b_1}$ as the size of the base bitrate segment. A bitrate $b^\uparrow$ denotes the next higher bitrate relative to $b$ in the set of discrete bitrates used to encode the video.

Each entry in $\mu$ corresponds with a likely behaviour of the consumer relative to the \textit{Ripple Bitrate}. The first case triggers when the requested bitrate matches the target rate for the router $b = RB^i$. In this case there are sufficient resources to satisfy subsequent requests at the requested bitrate. The reward function returns $\mu(b)$.

The second case is left for discussion following third and fourth cases. The third case returns when $b > RB^i$, the requested bitrate is higher than \textit{Ripple Bitrate}. Here a reward lower than $\mu(b)$ should be granted since the cache hit generates load or throughput that may cause consumers to reduce their video quality. As a result the lower reward discourages those cache partitions that can lead to video quality degradation. %\todo{Is next sentence needed? Appears to distract from my perspective, as well as being somewhat implicit.} The cache placement that maximizes this cache reward would be guided to satisfy requests as early as possible, since the reward value $\mu(RB^i)$ would also decrease along forwarding path.

The final case triggers when a cache is unable to maintain even the base rate video quality. In this case the $\gamma$ function returns the lowest reward of $\mu(b_1)$, since requests satisfied under such circumstances are likely to lead to buffer-induced freezing, and should be avoided.

Returning to the second case $b < RB^i$, when the requested rate is lower than \textit{Ripple Bitrate} for the cache. Recall that cache reward is only granted when there is a cache hit. Thus, this case represents a request that is satisfied by the cache, yet for content that should be pushed towards the network core. In this case the return value represents a trade-off. Strictly speaking, a cache hit encourages consumer to subsequently request a higher bitrate $b^\uparrow$. However, the additional load on the network could lead to bandwidth fluctuations that cause bitrate oscillation. Thus, care must be taken to avoid over-awards. $\gamma$ returns a weighted sum of $\mu(b)$ and $\mu(b^\uparrow)$, where the contribution of each component is controlled by parameter $\beta(b) \in [0, 1]$. $\beta(b) = 1$ returns $\gamma(RB^i,b) = \mu(b^\uparrow)$ and prioritizes video quality that consumers can achieve, while ignoring the risk of bitrate oscillation. Conversely, $\beta(b) = 0$ prioritizes bitrate stability by returning $ \mu(b)$. As \textit{RippleClassic} optimizes cache placement from a system-wide perspective, $\beta(b) = 0$ encourages \textit{RippleClassic} to relocate video content via matching \textit{Ripple Bitrate}. As a result, lower quality would be eventually pushed towards the network core.

The question then emerges: What is an appropriate weight? A fixed $\beta$ fails to capture the disproportional resource increases needed to satisfy requests for higher quality content. We then study users' QoE under a variable $\beta$, to highlight the trade-off under different design choices.
\begin{figure*}[!t]
	\centering
	\subfloat[Expected bitrate.]
	    {\includegraphics[width=0.3\textwidth]{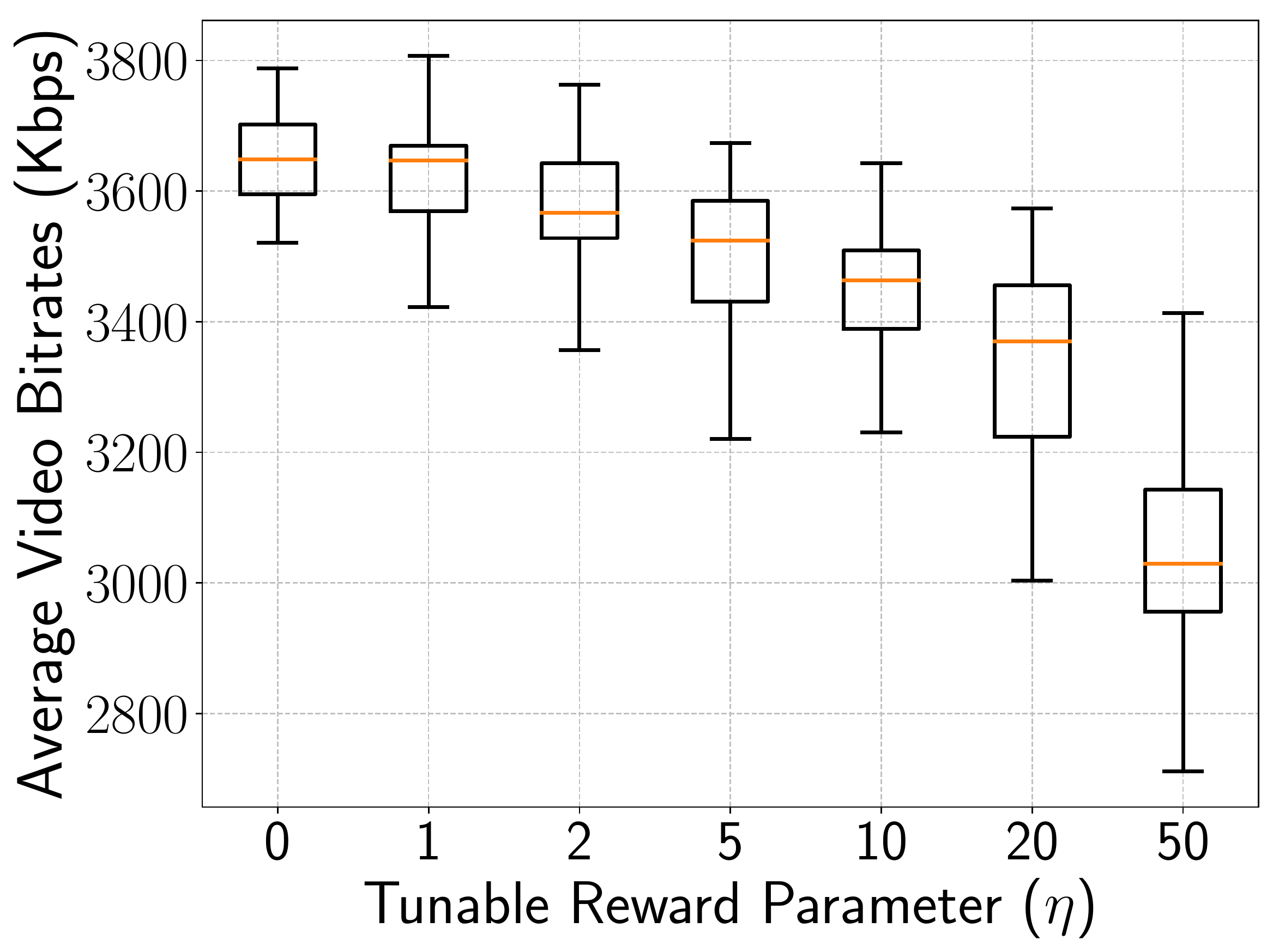}
	    \label{Fig:RewardSize}}
	\hfil
	\subfloat[Quality degradation (lower is better).]
	    {\includegraphics[width=0.3\textwidth]{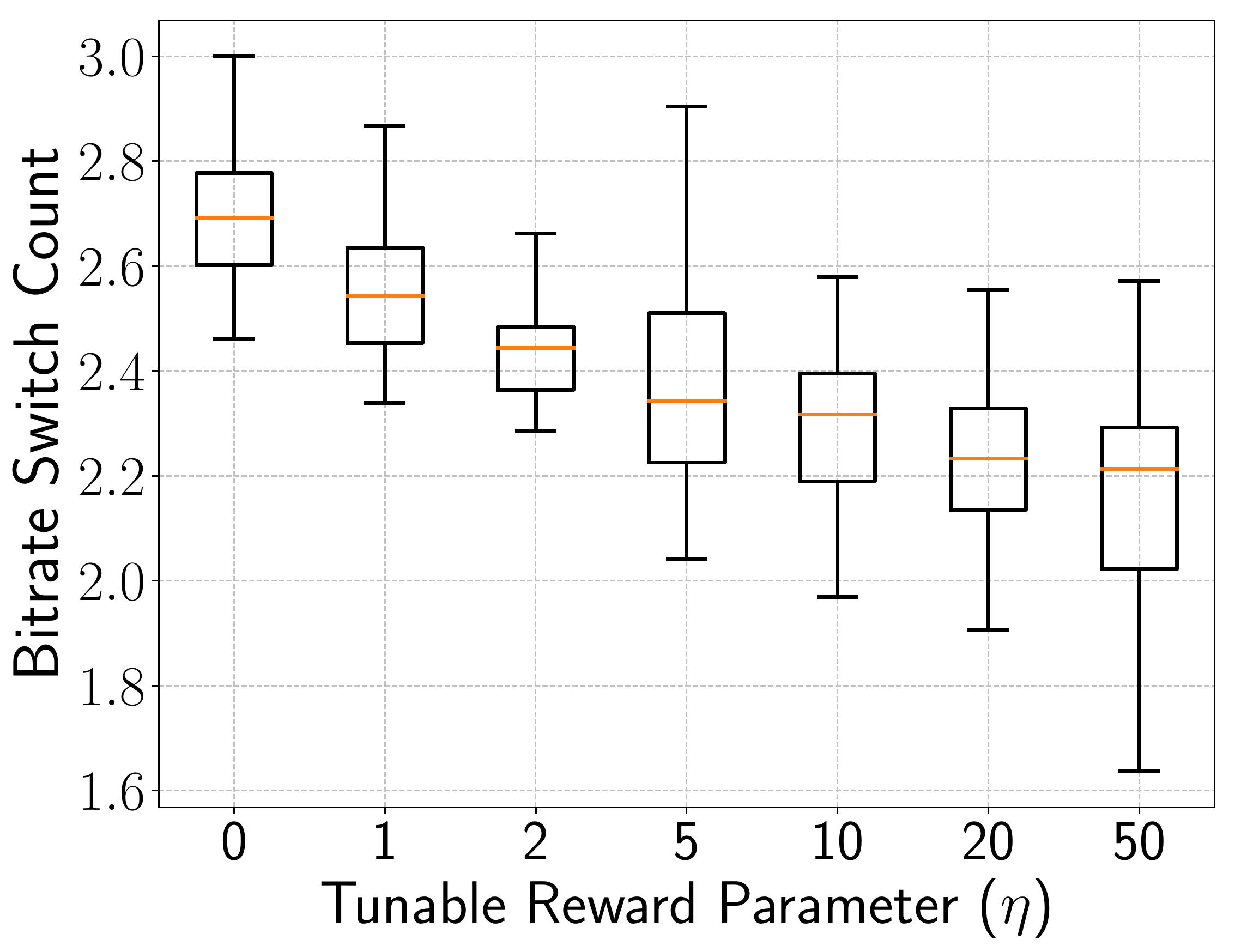}
		\label{Fig:RewardSwitch}}
	\hfil
	\subfloat[Buffer-induced freezing.]
	    {\includegraphics[width=0.3\textwidth]{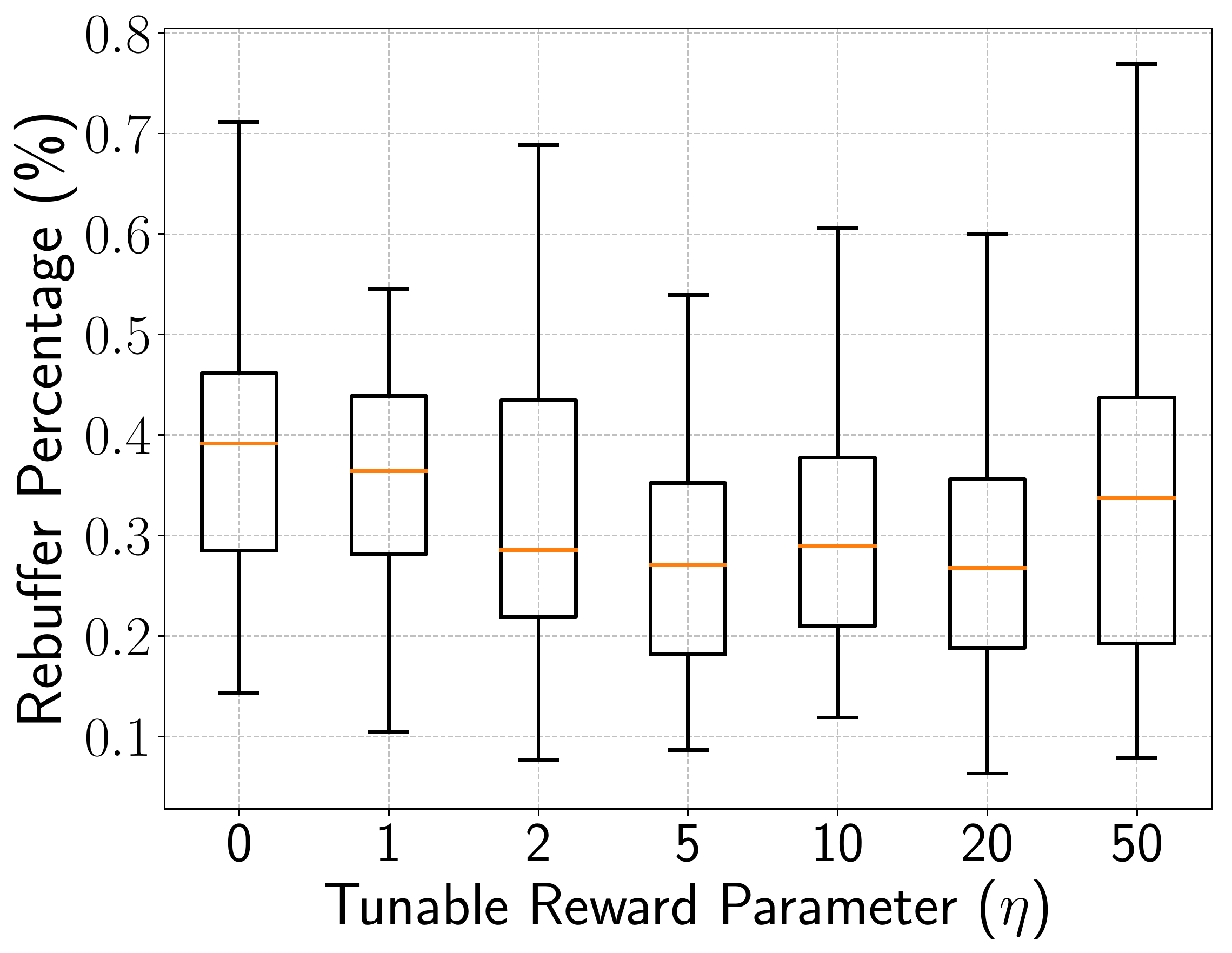}
		\label{Fig:RewardFreeze}}
	\caption{The impact of tunable cache reward $\eta$ on users' QoE.}
	\label{Fig:ImpactReward}
\end{figure*}

\subsection{Tuning the Quality-Oscillation Tradeoff}\label{subsec:eta}
We define $\beta(b)$ in a manner that is inversely proportional to the rank of the bitrate, $rank(b)$, such that
\begin{equation}
\beta(b) = \frac{1}{\eta + rank(b)}.
\end{equation}
The inverse of $rank(b)$ echoes the increasingly conservative nature of rate adaptation controls at higher quality, corresponding with the disproportional increases in resources to support higher bitrates. The high correlation revealed in our previous study between $\gamma$ rewards and consumer QoE implemented the inverse of $rank(b)$, alone~\cite{ourletter}. Here we add a tunable parameter $\eta$ to further explore the trade-off between quality and oscillation implied by $\beta$.

The competing demands between high quality and low oscillation are made evident by the box plots in Figure~\ref{Fig:ImpactReward}. These plots show the impact of $\beta$ on various measures of users' QoE for a range of $\eta$ values. Performance metric definitions, as well as further experimental design details, are provided in Section~\ref{Sec:Performance}.   
A smaller $\eta$ value favours $\mu(b^\uparrow)$, the reward that emphasizes higher quality. This can be seen in Figure~\ref{Fig:RewardSize}, where consumers receive the highest quality when $\eta=0$ and diminishing quality as $\eta$ values increase. Conversely, Figure~\ref{Fig:RewardSwitch} shows that those larger values of $\eta$ correspond with fewer adaptations that reduce quality. This happens because larger $\eta$ values emphasize stability via $\mu(b)$. Finally, Figure~\ref{Fig:RewardFreeze} shows no significant difference in buffer-induced freezing. We take this as evidence that consumer-side adaptations are able to ensure the same degree of uninterrupted playback despite changes in network conditions.

Figure~\ref{Fig:ImpactReward} points to $\eta=1$ as striking a good balance between bitrate and oscillation. As $\eta=0$ emphasizes video quality regardless of cache utilization and resulting bandwidth fluctuation, \textbf{$\eta=1$ would reduce bitrate oscillation without sacrificing on received video quality}. We found this to be true throughout our wider evaluations in Section~\ref{Sec:Performance}.

\textit{RippleClassic} is designed to be a benchmark partitioning scheme that optimizes for high-bitrate content by pushing lower-bitrate content towards the core. The complexity of \textit{RippleClassic} presents scalability challenges. In the next section, we design a distributed heuristic that can partition caches according to the \textit{RippleCache} principles in polynomial time.

\section{\textit{RippleFinder} Cache Partitioning}\label{Sec:RippleFinder}
The \textit{NP-Complete} complexity class of \textit{RippleClassic} is a barrier to deployment at scale. For larger networks, we design the distributed \textit{RippleFinder} cache placement scheme. \textit{RippleFinder} manages cache capacity per-forwarding path, rather than per-router. We begin with a high-level description, then follow with the details of each step, before showing that \textit{RippleFinder} executes in polynomial time.

\subsection{System Overview}
\begin{figure}[!b]
	\centering
	\includegraphics[width=\columnwidth]{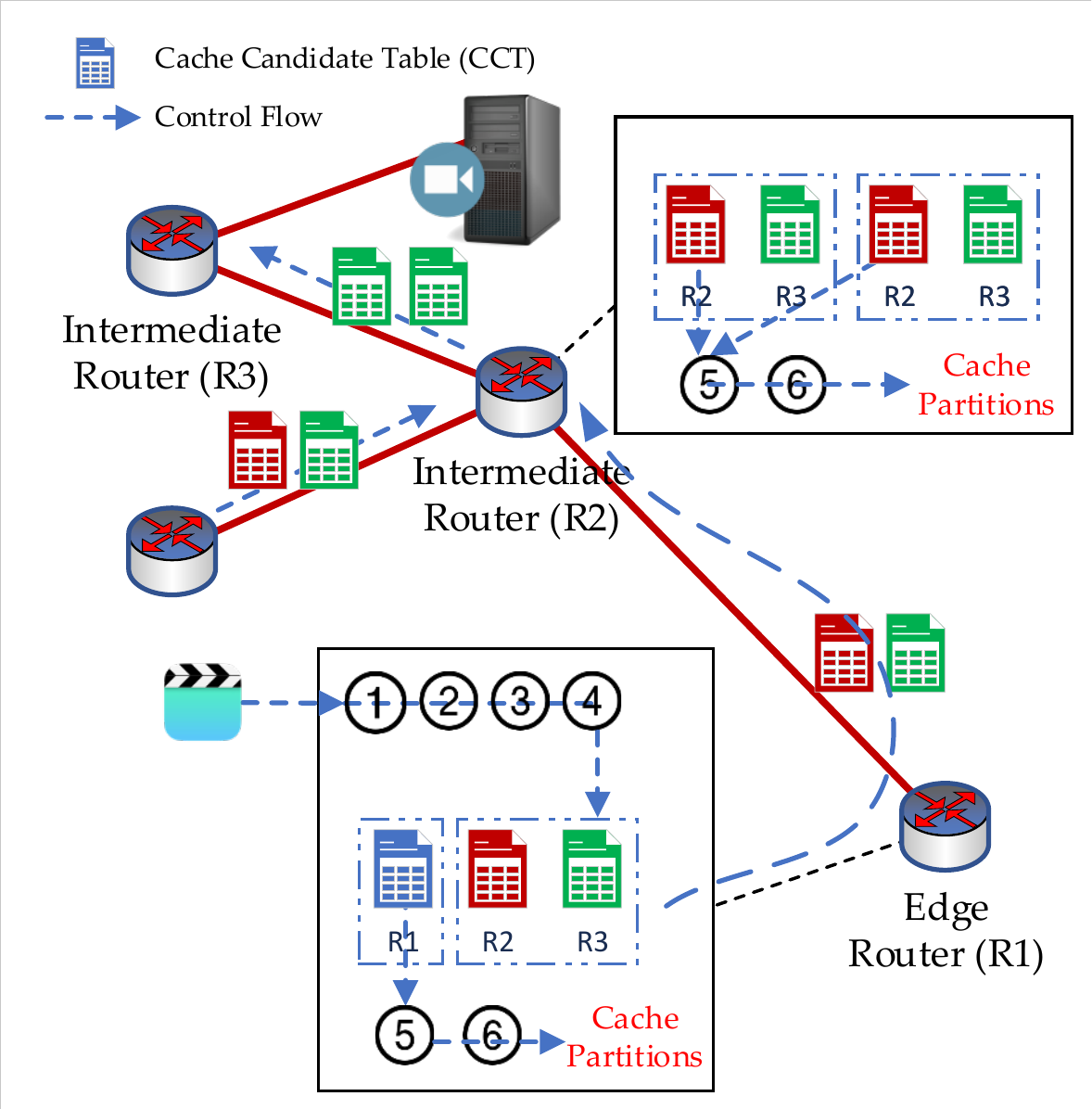}
	\caption{Execution of \textit{RippleFinder}. Edge router $R_1$ would create \textit{Cache Candidate Table}s (CCTs) for $R_1$, $R_2$ and $R_3$. CCT for $R_1$ is processed immediately on $R_1$. CCTs for $R_2$ and $R_3$ are delivered upstream. The intermediate router $R_2$ would intercept all CCTs for $R_2$ (the icon in red color), and forward CCTs for $R_3$.} \label{Fig:FinderSys}
\end{figure}

A full execution consists of 6 procedures performed in sequence. \textit{RippleFinder} begins at each edge router that (1)~\textit{ranks video segments} by their utility, and also (2)~\textit{discovers the total cache capacity} of the path. The edge router uses this information to (3)~\textit{push and pop} entries from the full ranking tables into new bitrate-specific stacks. Edge routers' final step is to (4)~\textit{nominate the caching candidates} for video content at each router on the forwarding path.

A system-wide representation appears in Figure~\ref{Fig:FinderSys}, where \textit{Cache Candidate Tables (CCTs)} for routers $R_1$, $R_2$ and $R_3$ (in blue, red, and green, respectively) are generated following Steps (1) (2) (3) and (4). $R_1$ is processed immediately on $R_1$, while the other two tables are delivered to upstream routers. Procedures (1)-(4) are repeated at each ICN edge router for each forwarding path.

Subsequent steps (5)-(6) are executed by all routers in the system. Routers that sit on intersecting paths must then (5)~\textit{negotiate their finite cache capacity} between competing paths once all candidate tables are received, since the total size of video segments in these candidate tables may exceed the cache capacity of this router. Note that the resulting cache capacity allocated to each path will differ from the initial values in Step~(2). In the final Step~(6), each router updates and returns this new values to the respective edge routers. 

Steps~(2) to~(6) are repeated until cache capacity values at nomination phase  (Step~(4)) match the values after negotiation (Step~(5)). This iteration is guaranteed to terminate, as is explained following the details of individual steps. 

\subsection{\textit{RippleFinder} in Execution}
Each individual step is described below with numbering that corresponds to the system overview.

\textbf{(1)~Ranking Table Construction}. Video statistics are used to rank content by utility, for each bitrate, as shown in Figure~\ref{Fig:RankingTable}. Every entry in a ranking table consists of the name of the content and the corresponding caching utility $\mathbb{U}$, sorted from high utility to low. The cache utility for video segment indexed by $(f,k,b)$ is calculated as,
\begin{equation}\label{Utility}
	\mathbb{U}_d(f,k,b) = \mu(b) * \theta_d(f,k,b).
\end{equation}
$\mu(b)$ and $\theta_d(f,k,b)$, both previously defined in Section~\ref{Sec:Optimization}, are a value proportional to the size of video segment and the number of requests, respectively. This notion of utility emphasizes video content that is both costly to deliver and highly popular. The caching decisions would then cater to video segments with high overall utility. 

\textbf{(2)~Cache Capacity Discovery}. The core of \textit{RippleFinder} manages the entire cache capacity along each forwarding path. In this step, available cache volumes of routers dedicated to a forwarding path $[d,p]$ of length $L$ are concatenated so that the total path capacity is $C_{[d,p]} = \sum_{j = 1}^L C_{[d,p]}^j$. We note that $C_{[d,p]}^j$ differs from our earlier definition of $C_v$, where $C_v$ represents the entire caching space on a certain router $v$. For any $j = v, C_{[d,p]}^j \leq C_v$ since the volume at a router dedicated to a path must be upper bounded by the router's cache capacity. In \textit{RippleFinder}, the initial value for $C_{[d,p]}^j \leftarrow C_v$. However, as caching decisions are made along each forwarding path independently and routers in an ICN may be shared by multiple paths, one cannot guarantee that our initial assumption always remains valid. As shown in Figure~\ref{Fig:DiscoverCapacity}, only portions of the cache capacity at ICN nodes may be allocated to a forwarding path, so that some capacity may be reserved to content delivered through other paths. 
%While it may be that edge routers can manipulate the caching space of all routers on the path, RippleFinder makes no such assumption. 
The volume of a cache on the path may be adjusted in later steps. Consequently, the cache capacity discovery marks the beginning of an iteration that ends with cache volume updates in Step~(6).

\textbf{(3)~"Push" and "Pop"}. In this intermediate step, a cache stack $ST_b$ is populated for each of the bitrate ranking tables in Step~(1).
Entries from ranking tables are pushed into the corresponding stack in descending order. The ordering can be seen in Figure~\ref{Fig:PushAndPop}, where higher bitrate stacks are filled before lower bitrate stacks, and within each stack the higher utility items sit deeper than lower utility items. Ranked entries are pushed into the stacks until the `stacked' size of the video segments exceeds the total available cache capacity, i.e.\@ 
\begin{equation}\label{Constraint}
	\sum_{b\in B} Size(ST_b) > C_{[d,p]}.
\end{equation}
Once the cache size required by stack elements exceeds capacity $C_{[d,p]}$, RippleFinder pops and pushes entries as follows, and depicted by example in Figure~\ref{Fig:PushAndPop}. Until constraint $\sum_{b\in B} Size(ST_b) \leq C_{[d,p]}$ is restored, \textit{RippleFinder} compares the top entries of each stack and pops the entry with lowest utility. Note that it is possible for a least-utility entry in a high-bitrate stack to have less utility than the least-utility entry in a lower-bitrate stacks. Once constraint $\sum_{b\in B} Size(ST_b) \leq C_{[d,p]}$ is restored, pushing resumes as normal until the known capacity is again exceeded. Stack operations continue until the lowest bitrate stack is marked \textit{complete}. A stack is marked complete when the popped video content is taken from the stack that is currently being filled since the overall cache utility can no longer be improved by continuing to push content into the current stack. The ordering of push operations ensures that higher bitrate stacks will always be marked complete before lower quality stacks. The content corresponding to entries that have been popped or that remain in the ranking tables would be excluded from cache placement.
\begin{figure}[!t]
\renewcommand*\thesubfigure{\arabic{subfigure}}
\makeatletter
\renewcommand{\p@subfigure}{\thefigure-}
\makeatother
		\centering
		\subfloat[\textbf{Ranking Table Construction}. Construct ranking tables for each bitrate ($B_1$, $B_2$ and $B_3$) from video statistics collected by edge router.]{\includegraphics[width=\columnwidth]{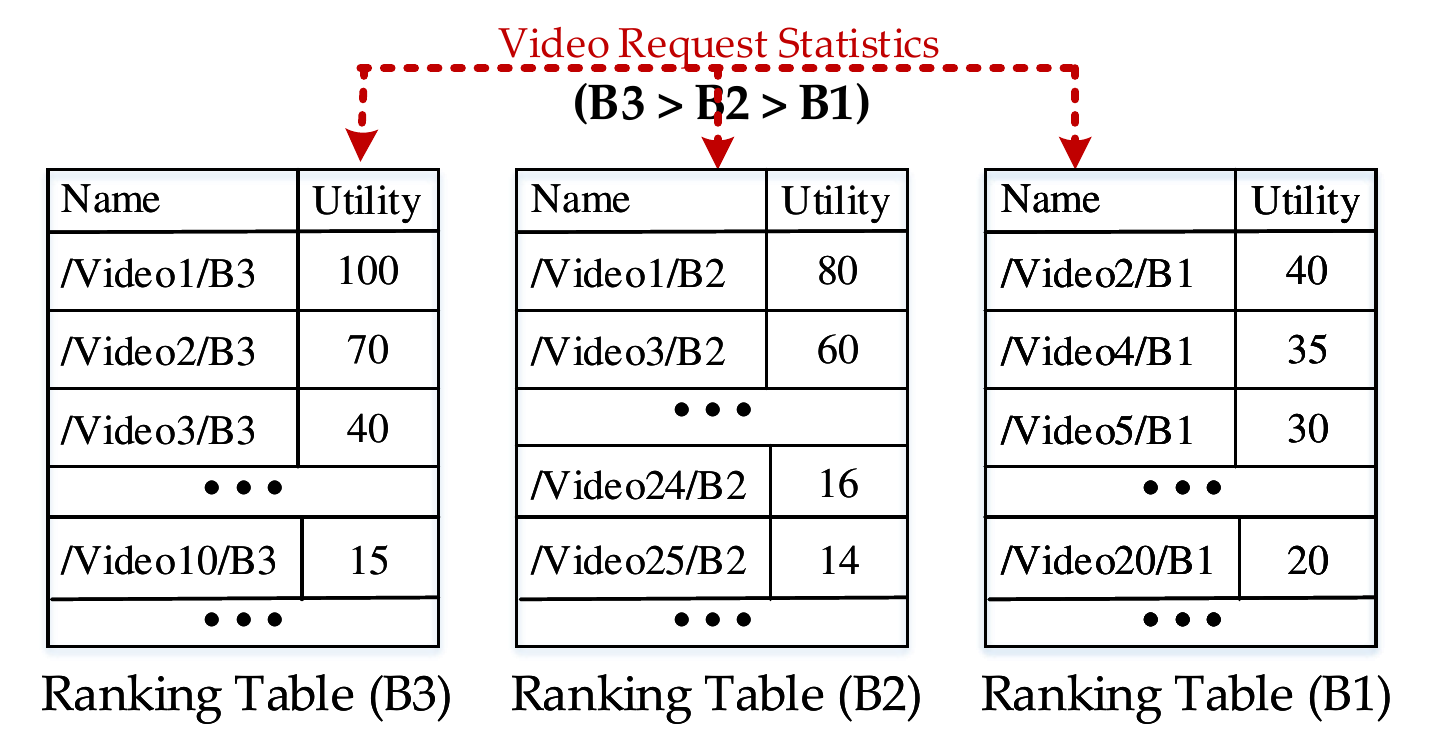}
			\label{Fig:RankingTable}}
		\vspace{2em}
		\subfloat[\textbf{Cache Capacity Discovery}. The shaded volume of router $R_1$ , $R_2$ and $R_3$ would be used to cache video content delivered along forwarding path. These shaded volume is added together, with a size of $C_{[d,p]}$.]{\includegraphics[width=\columnwidth]{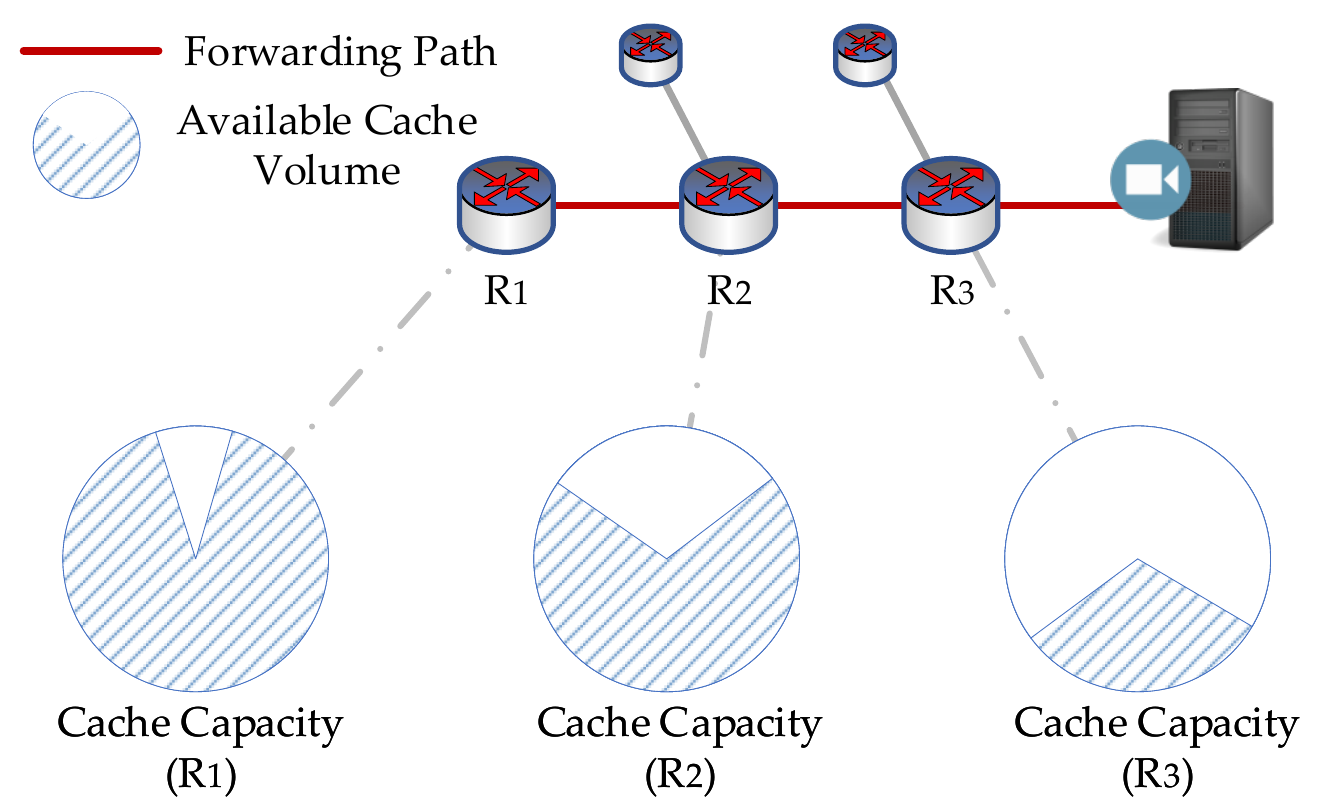}
			\label{Fig:DiscoverCapacity}}
		\vspace{2em}
		\subfloat[\textbf{``Push'' and ``Pop''}. Video content is pushed into \textit{Cache Stack} by ranking order. After content `\textit{/Video25/B2}' is pushed into $ST_{B_2}$, the Equation~\ref{Constraint} is violated, which triggers `Pop' operation. Since utility of `\textit{/Video10/B3}' on top of $ST_{B_3}$ is higher than `\textit{/Video25/B2}' on top of $ST_{B_2}$, video segment  `\textit{/Video25/B2}' is popped.]{\includegraphics[width=\columnwidth]{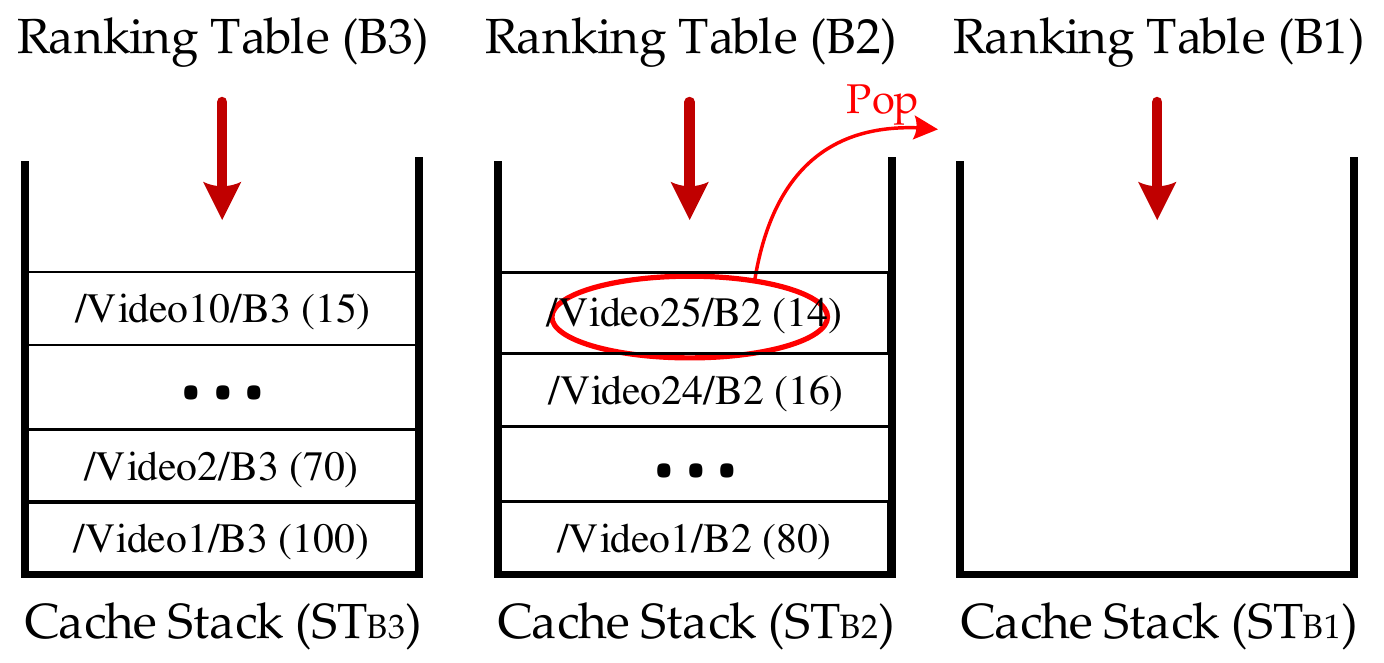}
			\label{Fig:PushAndPop}}
		\caption{\textit{RippleFinder} in execution.}
		\vspace{-0.5em}
\end{figure}
\begin{figure}[!t]
\renewcommand*\thesubfigure{\arabic{subfigure}}
\makeatletter
\renewcommand{\p@subfigure}{\thefigure-}
\makeatother
\ContinuedFloat
		\centering
		\subfloat[\textbf{Cache Candidate Nomination}. Video segments in \textit{Cache Stack}s are assigned to \textit{Cache Candidate Table}s (CCTs). The assignment occurs first at CCT for $R_1$, followed by $R_2$ and $R_3$. Video content in $ST_{B_3}$ is first arranged, followed by $ST_{B_2}$ and $ST_{B_1}$.]{\includegraphics[width=\columnwidth]{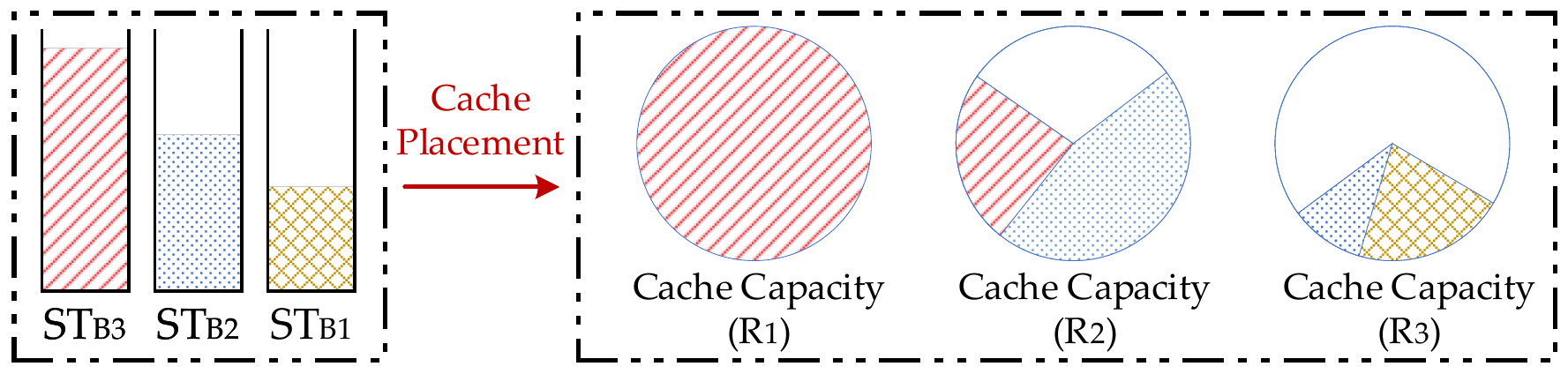}
		\label{Fig:Nominate}}
		\vspace{2em}
		\subfloat[\textbf{Cache Placement Negotiation}. The cache placement decision is made by (1) merging CCTs received from path ${[d,p]}$ and path ${[d',p']}$, and (2) choosing video segments with high utility as long as the cache capacity of this router $C_v$ allows. ]{\includegraphics[width=\columnwidth]{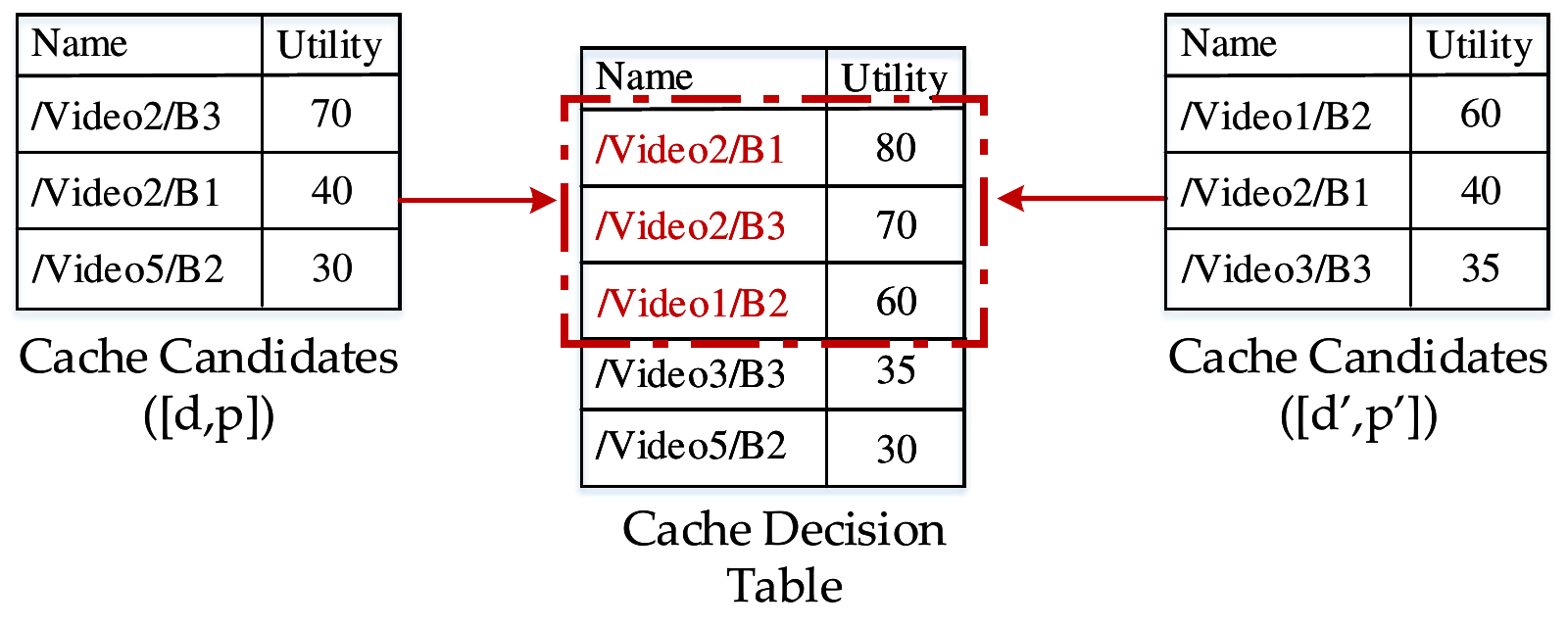}
			\label{Fig:Negotiation}}
		\vspace{2em}
		\subfloat[\textbf{Cache Volume Update}. The final caching decisions are compared against the items in each CCT. Segments `\textit{/Video2/B1}' and `\textit{/Video2/B3}' appear in both final cache placement and CCT. The updated cache volume of this router to path ${[d,p]}$ would be equal to the total size of these two segments. ]{\includegraphics[width=\columnwidth]{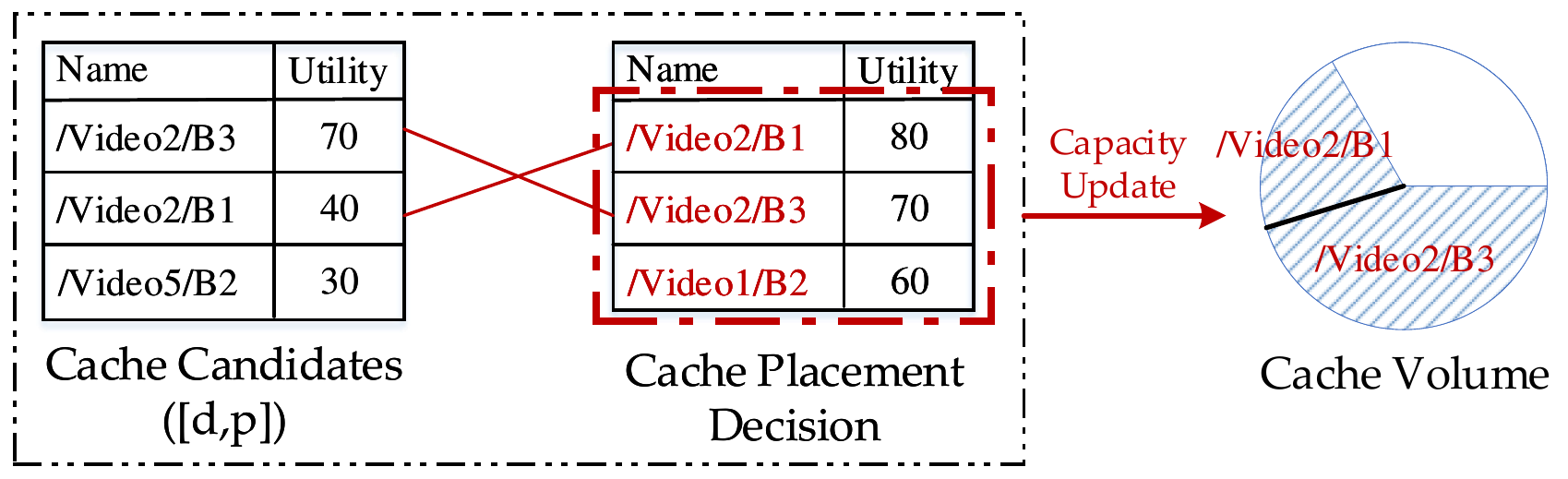}
			\label{Fig:VolumeUpdate}}
		\caption{\textit{RippleFinder} in execution.}
		\vspace{-0.5em}
\end{figure}

\textbf{(4)~Cache Candidate Nomination}. For each cache node along the forwarding path, the edge router constructs a Cache Candidate Table (CCT). Tables are populated with entries from the cache stacks, again in descending bitrate order, starting from the CCT for edge router itself. We note that content from any stack may span multiple tables. For example, the depiction in Figure~\ref{Fig:Nominate} shows content from the stack for ${B_3}$ has filled the CCT for router $R_1$, and overflows into the CCT for $R_2$. The sum sizes of content assigned to candidate tables are capped by the capacities reported to the edge router during discovery in Step~(2). Since the total space required by all items in all stacks is constrained by the path capacity $C_{[d,p]}$, every item in the stacks finds a place in a CCT.
%\todo{Strictly speaking, without some kind of `rounding' maybe(?), this is NOT true because of the discrete granularity of segments sizes vs bytes on disk}
The result adheres to \textit{RippleCache} ideals by assigning high-bitrate content in tables for ICN routers closer to consumers, leaving lower bitrate content for CCTs bound towards the core. 

\textbf{(5)~Cache Placement Negotiation}. Cache nodes receive a candidate table from each of its forwarding paths. The combined entries from all CCTs may exceed the cache's capacity and must be negotiated. Nodes 
rank video segments from all CCTs by the \textit{sum} of the content's utility. In Figure~\ref{Fig:Negotiation} the individual utility values for `\textit{/Video2/B1}' from left and right tables are summed to a utility of $80$. Each router $v \in \mathbb{V}$ can cache up to its cache capacity $C_v$, according to the sorted utility from high to low.

\textbf{(6)~Cache Volume Update}. Since sum utility is used to populate a cache, the portion of capacity dedicated to a path may be smaller than was previously reported in Step~(2). 
%This may be the consequence of relative utility, where a volume for one path shrinks to accommodate a volume with higher utility content from another path. Multiple volumes in a cache may also adjust simultaneously when content is shared between paths. 
In the example shown by Figure~\ref{Fig:VolumeUpdate}, a router would cache the three video segments enveloped in the red dot-dash line. The volume size dedicated to path $[d,p]$ is then equal to the size of both segments `\textit{/Video2/B3}' and `\textit{/Video2/B1}'. 
Updated volume sizes are returned to the respective edge routers along the reverse of forwarding paths, so that the available cache capacity $C_{[d,p]}$ for a entire path can remain current.

At this stage, the updated volume sizes are compared with previous values obtained in Step~(2). A mismatch triggers another iteration of Steps~(2)-(6). \textit{RippleFinder} terminates when $C_{[d,p]}$ is unchanged for all forwarding paths between two consecutive iterations. \textit{RippleFinder} is guaranteed to terminate. In any iteration, candidate tables are constructed with reported cache volume sizes. Since cache candidates nominated by edge routers may be omitted from the final placement at core routers, $C_{[d,p]}^j$ decreases monotonically. The worst possible case is that no cache capacity is allocated for a path, meaning that $C_{[d,p]}^j$ is capped by 0. Since no volume can be negative iteration must eventually end.

\subsection{\textit{RippleFinder} algorithm and complexity}
\textit{RippleFinder} is a distributed algorithm with polynomial time complexity of the number of paths in the system, $|\mathbb{D}|\cdot|\mathbb{P}|$. 
For ease of presentation, \textit{RippleFinder} is written as a single-thread of execution in Algorithm~\ref{Alg:RippleFinder}. 
\begin{algorithm}[H]
	\caption{\emph{RippleFinder}}
	\label{Alg:RippleFinder}
	\algsetup{
		linenodelimiter=:
		%indent=2em
	}
	\begin{algorithmic}[1]
		\renewcommand{\algorithmicrequire}{\textbf{Input:}}
		\renewcommand{\algorithmicensure}{\textbf{Output:}}
		\renewcommand{\algorithmiccomment}[1]{\\// #1}
		\REQUIRE Edge router $d$; Set of Producers $\mathbb{P}$; Length of routing path $L$ for each $(d,p), p\in\mathbb{P}$; Dedicated cache volume $C_{[d,p]}^j$ at each hop.
		\ENSURE  Adaptation-aware cache placement $x_d$ on router $d$.
		
		\STATE{Initialize available cache volume $C_{[d,p]} \leftarrow \sum_{j=1}^L C_{[d,p]}^j$}
		\REPEAT
		\FORALL{$p \in \mathbb{P}$}
				\STATE \textit{Ranking Table Construction}
				\STATE $C_{[d,p]} \leftarrow$ \textit{Cache Capacity Discovery}
				\STATE \textit{``Push'' and ``Pop''}
				\STATE $CCT \leftarrow$ \textit{Cache Candidate Nomination}
		\ENDFOR
		\COMMENT{$j=1$ as \textit{RippleFinder} is working on an edge node.}
		\STATE $C_{[d,p]}^1, x_d \leftarrow$ \textit{Cache Placement Negotiation}
		\FORALL{$p \in \mathbb{P}$}
		\STATE{$C_{[d,p]}' \leftarrow$ \textit{Cache Volume Update}}
		\ENDFOR
		\UNTIL{$C_{[d,p]} = C_{[d,p]}'$}
		\RETURN $x_d$.
	\end{algorithmic}
\end{algorithm}
The analysis pertains to edge routers since only edge routers execute the full set of operations; intermediate routers are limited to negotiating placements and updating cache volumes. Line 4 constructs $B$ ranking tables, each of up to size $FK$, with sorting complexity $\mathcal{O}(B\cdot FK\log{(FK)})$. Line 5 iterates over every router in each forwarding path to update cache capacity, with a complexity of $\mathcal{O}(|\mathbb{V}|)$. Both Lines 6 and 7 each scan over existing data structures in a time that is linear with the size of the structures, of $\mathcal{O}(BFK)$.  Thus, the overall complexity for the full set of tasks (Line 3-8) is $\mathcal{O}(|\mathbb{P}|BFK\log{(FK)} + |\mathbb{P}||\mathbb{V}|)$, 
as the same operations have to repeat for totally $|\mathbb{P}|$ number of forwarding paths. The complexity incurred by Steps~(5) and~(6) at all routers is dominated by merging CCTs at Line 9. CCT is already a sorted table, and the length of each CCT is capped by cache capacity and also expected to be markedly less than the length of ranking tables (that contain all requested video content) at Line 4. As such the complexity at Line 9 is bound by that at Line 4. Therefore, the complexity of \textit{RippleFinder} is $\mathcal{O}(|\mathbb{P}|BFK\log{(FK)} + |\mathbb{P}||\mathbb{V}|)$.

\section{Performance Results And Insights}\label{Sec:Performance}
We evaluate \emph{RippleClassic} and \textit{RippleFinder} performance via simulation against known caching strategies on the Named Data Networking (NDN) architecture. Results reinforce the broader merits of cache partitioning for adaptive streaming. We claim without loss of generality that the merits of \textit{RippleCache} designs and subsequent analyses can be applied on other ICN architectures and cache hierarchies.

\subsection{Simulation Setup and Parameters}
The proposed cache partitioning schemes were implemented onto ndnSIM~\cite{alexander2012ndnsim}, an NS-3 based simulator. Each NDN router is allocated a Content Store (CS), where its size $C_v$ is subject to a total available system capacity, controlled by $\omega$, as 
\begin{equation}
C_v = \frac{\sum \textnormal{Size of Video}}{\textnormal{\# of NDN Routers}} * \omega, \forall v \in \mathbb{V}.\nonumber
\end{equation}
Consumer-side adaptation behaviour is simulated via our own implementation of FESTIVE~\cite{jiang2012improving}, a throughput-based mechanism that captures recent advancements in bitrate adaptation. Users' interests in video content vary across different video files, captured by a \emph{Zipf}-like distribution (controlled via skewness parameter $\alpha$).  Videos are comprised of 4-second segments. Each video segment is prepared at 1, 2.5, 5, and 8 Mbps, which are recommended encoding bitrates by YouTube~\cite{youtubelink}.  Consumers initiate a session first by requesting a video file and retrieving video-related meta-data (i.e.\@ the Media Presentation Description (MPD)) from the producer. Interests in videos are triggered following a Poisson process, with an average time interval between two consecutive interests as 300 seconds. Once interest for a video file is initiated, subsequent requests for the session are initiated by the bitrate adaptation algorithm. 

Three additional caching schemes are evaluated alongside our proposed \textit{RippleFinder} and \textit{RippleClassic} for comparison. \emph{Cache Everything Everywhere (CE2)}~\cite{zhang2015survey} with LRU,  also with LFU, is a baseline that commonly appears in literature~\cite{zhang2015survey}. \emph{ProbCache}~\cite{psaras2012probabilistic} serves as a baseline for probabilistic caching~\cite{cho2012wave}. Both \textit{RippleClassic} and \textit{RippleFinder} are cache placement schemes. As the interaction between in-network caches and consumer-side adaptation exists, the caching decisions from \textit{RippleClassic} and \textit{RippleFinder} are updated iteratively to keep up with the changes on users' preferred bitrates. The iteration on \textit{RippleClassic} stops once two consecutive optimization produces similar cache rewards. \textit{RippleFinder} stops after a fixed number of iterations, where we observe a stable performance on users' QoE.

\begin{table}[!t]
	\centering
	\caption{Simulation Parameters}
	\label{Table:Parameters}
	\begin{tabular}{>{\quad}lll} \toprule
		\multicolumn{1}{l}{\textbf{NDN}} & BIP-tractable & Large-scale\\ \midrule
		Number of video files & 25 & 500\\
		Number of video segments per file & 25 & 50\\
		Number of NDN routers & 16 & 42\\
		Video segment playback time (sec) & 4 & 4\\
		Number of video producers & 1 & 3\\
		Number of video consumers & 32 & 84\\
		Encoded bitrates (Mbps) &\{1, 2.5, 5, 8\} &\{1, 2.5, 5, 8\}\\
		Request interval on video file (sec) & 300 & 300\\
		Bandwidth (Mbps) & 20 & 20\\
		Cache reward parameter ($\eta$) & 1 &\\
		Skewness factor ($\alpha$) & 1.2 & 1.2\\
		Content store size percentage ($\omega$) & 0.2 & 0.05\\ \midrule
		\multicolumn{3}{l}{\textbf{FESTIVE}}\\ \midrule
		Drop Threshold & 0.8 & 0.8\\
		Combine Weight & 8 & 8\\
		\bottomrule\end{tabular}
\end{table}

Two separate networks are implemented for evaluation. A smaller 16-node ICN network with a maximum 7-hop distance from a video producer to consumers allows the binary integer programming from \textit{RippleClassic} to find solutions in reasonable time. Variations on hop distance are used to cause different video access delay by consumers. We choose network link capacity at 20 Mbps, and as a result, the highest bitrate (8 Mbps) cannot be retrieved directly from the producer and must be provided by caches. We choose this relatively small link capacity to examine the performance that is enhanced by caching policies. 
Results for \textit{RippleClassic} are shown for $\eta=1$. Recall from Figure~\ref{Fig:ImpactReward} and surrounding discussion that $\eta=1$ appears to strike a good balance between prioritizing high bitrates and low oscillation. 

A larger 42-node topology generated by BRITE~\cite{medina2001brite} is used to evaluate cache partitioning in a realistic and large-scale system with multiple producers. The complete list of simulation parameters for both scenarios/topologies are listed in Table~\ref{Table:Parameters}. 

Results are evaluated using standard QoE metrics published by the DASH Industry Forum~\cite{dashmetrics}. From the standard set, we adopt three metrics, \textit{Average Video Quality}, \textit{Bitrate Switch Count} and \textit{Rebuffer Percentage}, as described in their relevant sections.
% defined as follows:
% \begin{inparaenum}[(i)]
% 	\item \textit{Average Video Quality} is the average video quality that consumers request among all video sessions;
% 	\item the \textit{Bitrate Switch Count} tracks the number of times the requested video bitrate changed during a video session; and
% 	\item \textit{Rebuffer Percentage} is the average time spent in a video freezing state over the active time of a video session.
% \end{inparaenum}
Each set of evaluations is repeated across a range of content store size ratio $\omega$ and popularity-skewness parameter $\alpha$. All results are presented at a 95\% confidence level.

\begin{figure*}[!t]
	\centering
	\begin{minipage}{\textwidth}
	    \centering
	    \subfloat[Average Video Bitrate]
	    {\includegraphics[width=0.32\textwidth]{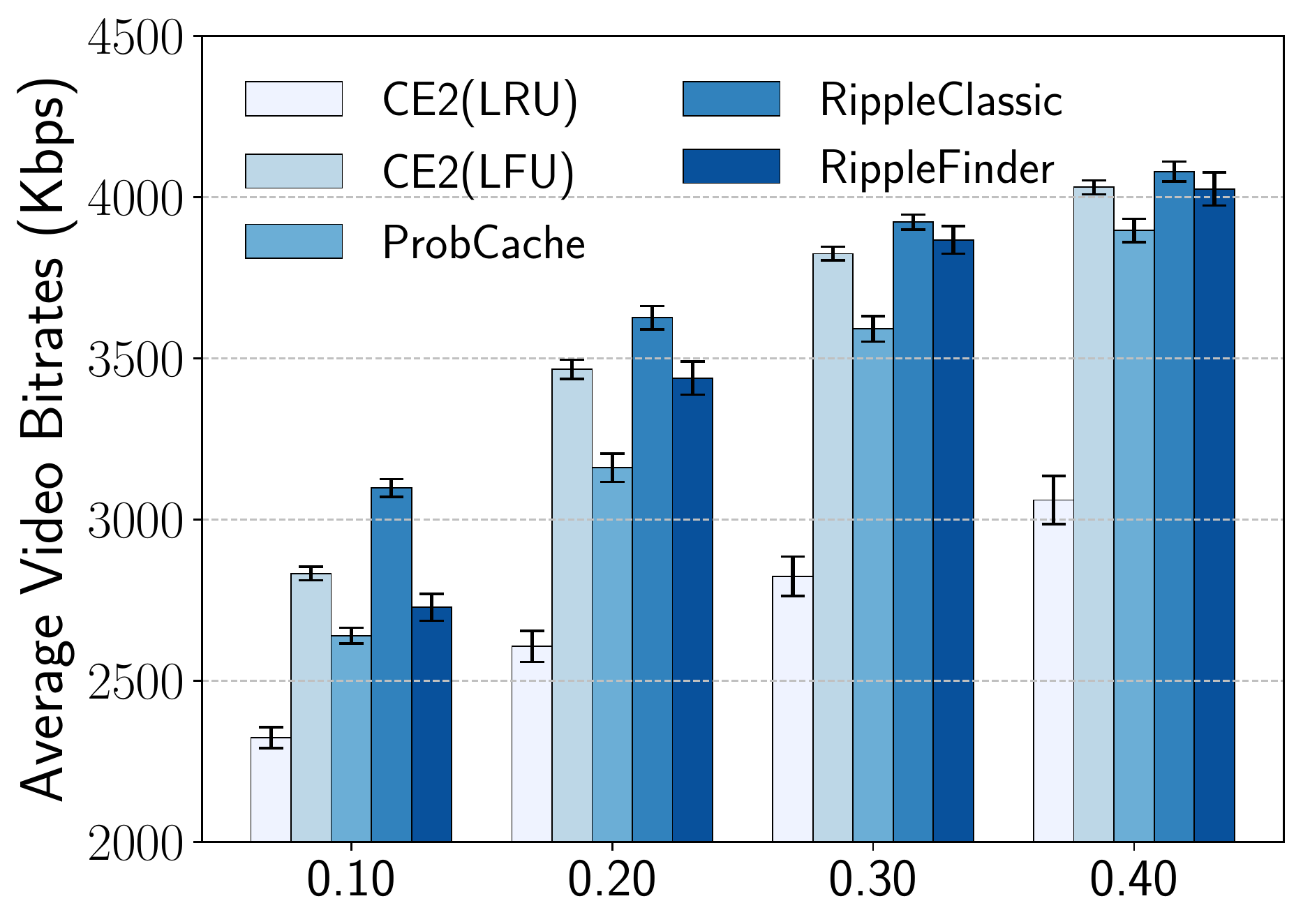}
		\label{Fig:BRSize}}
		\hfil
		\subfloat[Bitrate Switch Count]
		{\includegraphics[width=0.32\textwidth]{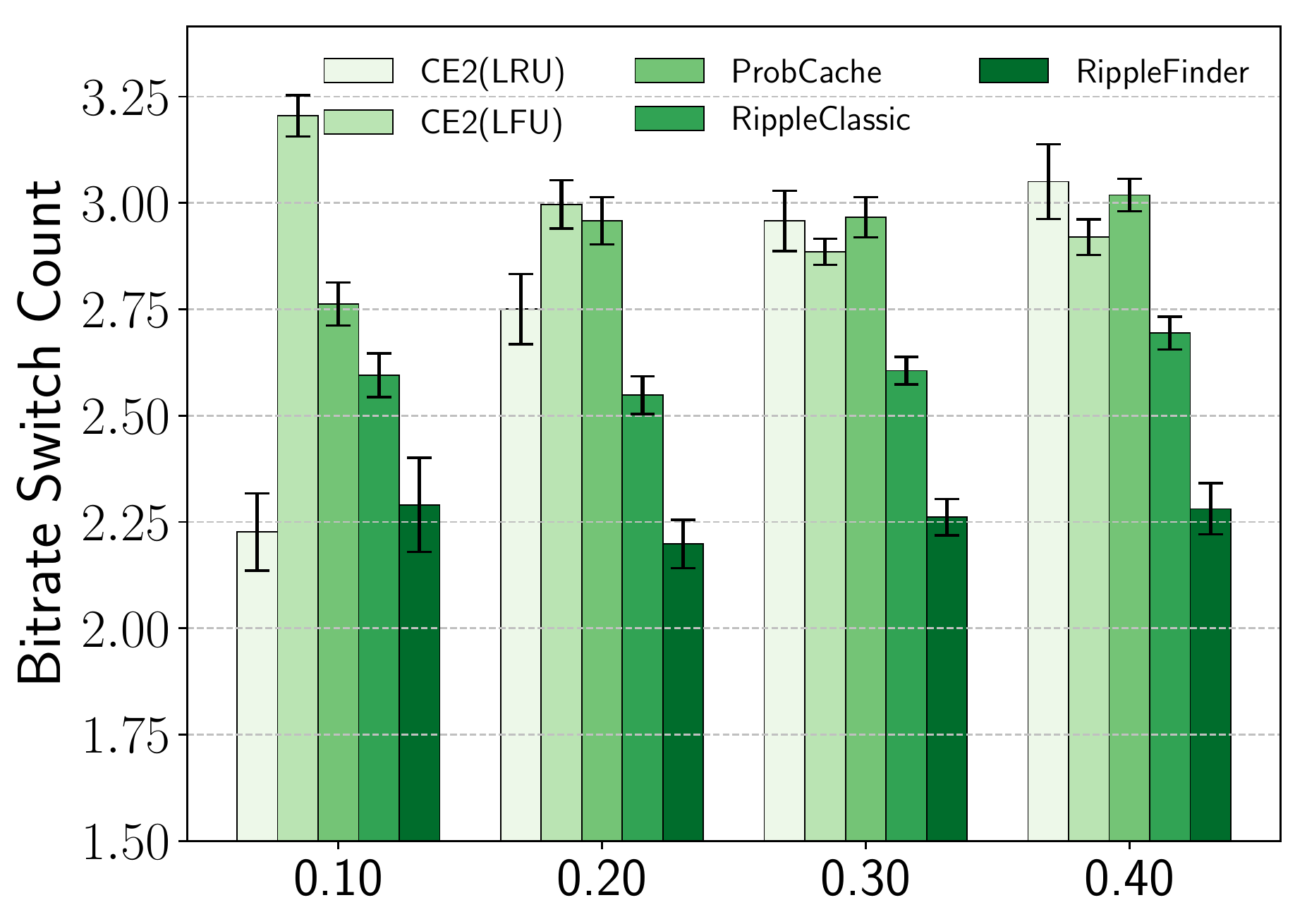}
		\label{Fig:BOSize}}
		\hfil
		\subfloat[Rebuffer Percentage]
		{\includegraphics[width=0.32\textwidth]{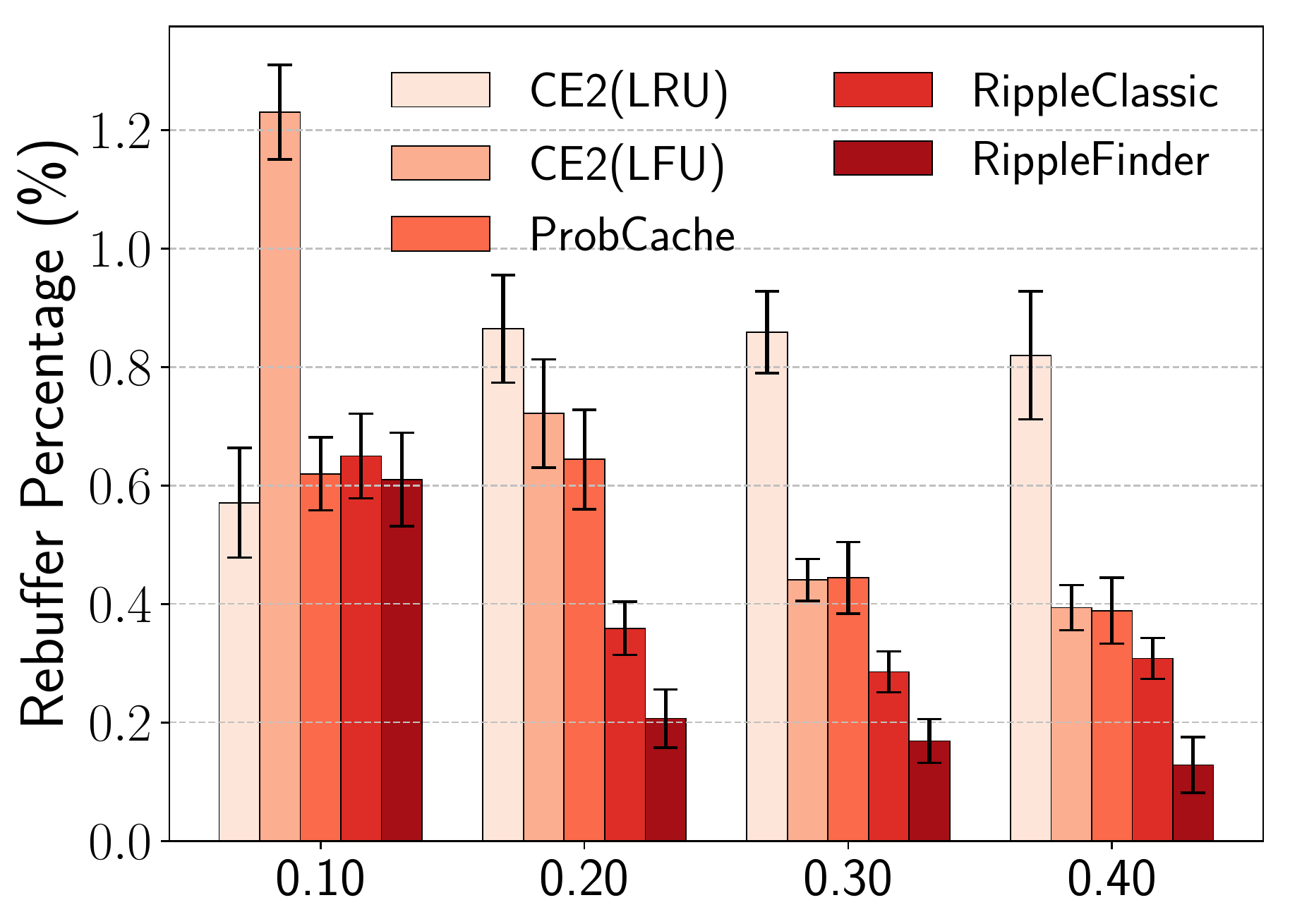}
		\label{Fig:FreezeSize}}
		\caption{Content Store Size Percentage ($\omega$) for `BIP-tractable' settings}
		\label{Fig:SizeEvals}
	\end{minipage}
	\begin{minipage}{\textwidth}
	    \centering
		\subfloat[Average Video Bitrate]
		{\includegraphics[width=0.32\textwidth]{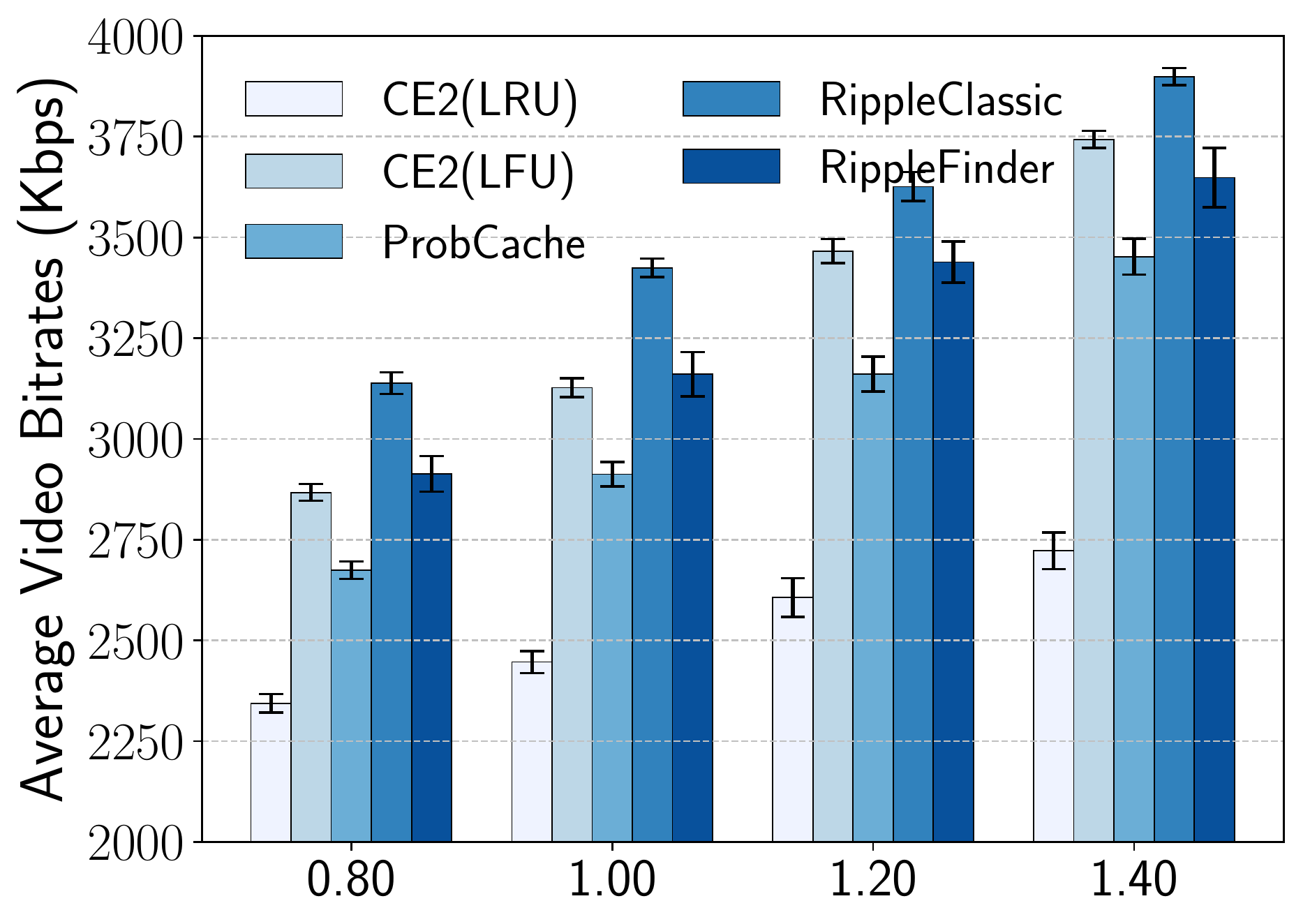}
		\label{Fig:BRZipf}}
		\hfil
		\subfloat[Bitrate Switch Count]
		{\includegraphics[width=0.32\textwidth]{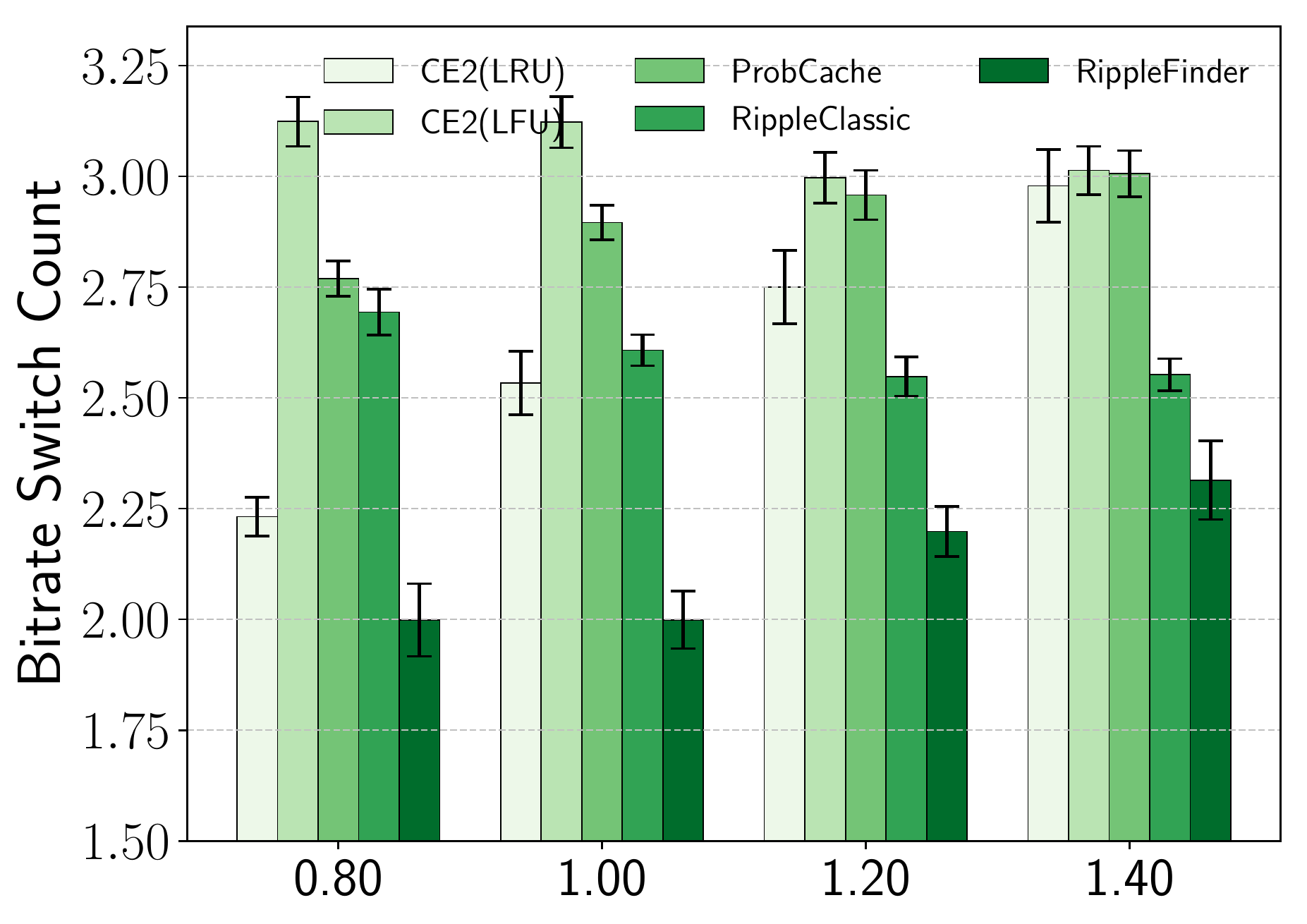}
		\label{Fig:BOZipf}}
		\hfil
		\subfloat[Rebuffer Percentage]
		{\includegraphics[width=0.32\textwidth]{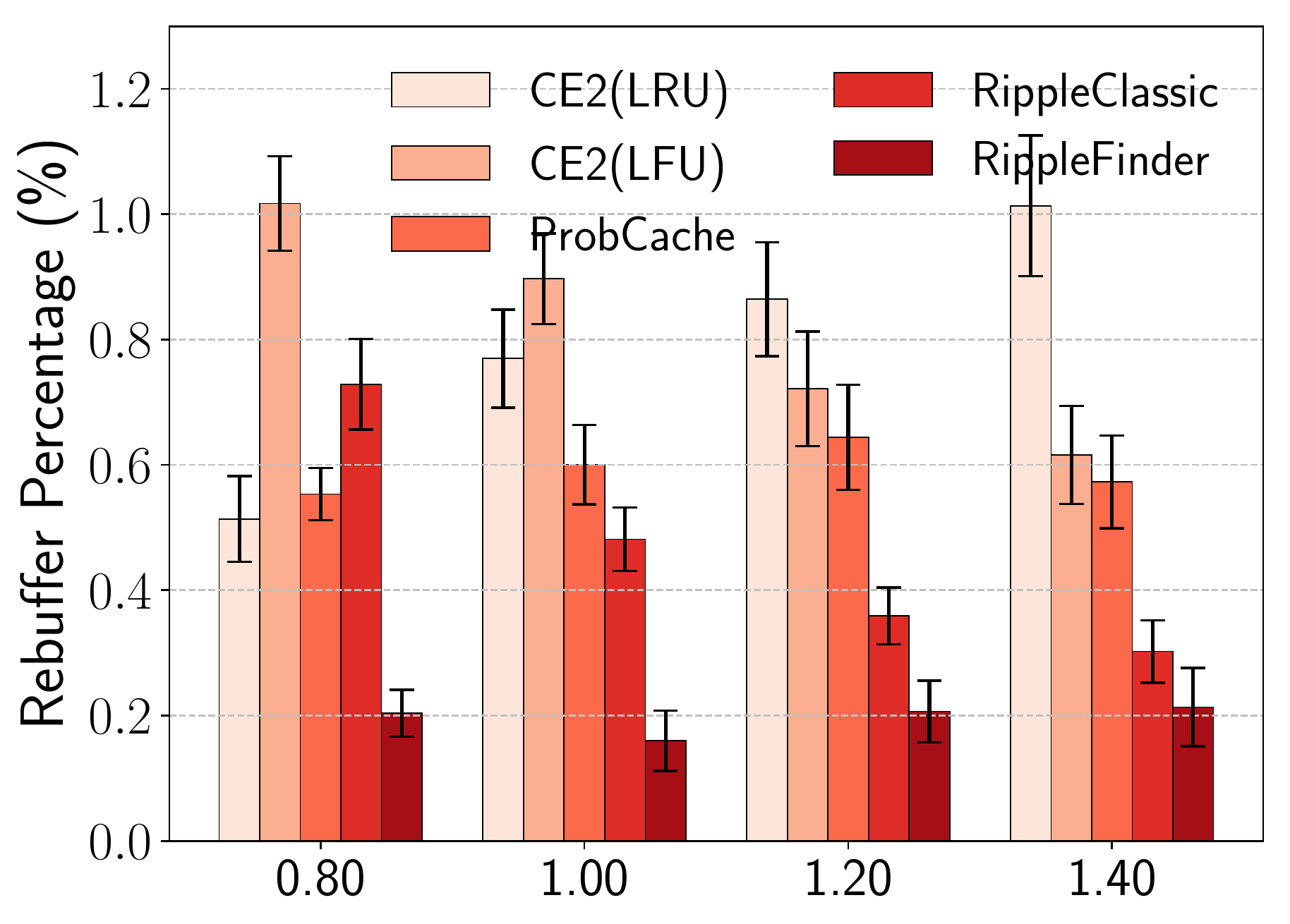}
		\label{Fig:FreezeZipf}}
        \caption{Popularity Skewness ($\alpha$) for `BIP-tractable' settings}
        \label{Fig:ZipfEvals}
    \end{minipage}
\end{figure*}

\subsection{Average Video Quality}\label{subsec:impactQuality}
Figures~\ref{Fig:BRSize} and~\ref{Fig:BRZipf} show the
\textit{Average Video Quality}, defined as the average video bitrate that consumers request among all video sessions~\cite{dashmetrics}. Measurements indicate that \textit{RippleClassic} and \textit{RippleFinder} performance meet or exceed \textit{CE2} with LFU and \textit{ProbCache}. We observe that the gap in performance against the benchmark \textit{RippleClassic} grows proportionally larger as cache resources diminish. For example, Figure~\ref{Fig:BRSize} shows that when the cache capacity ratio is $\omega = 0.2$, \textit{RippleClassic} delivers higher video quality than \textit{CE2} with LFU by 4.6\%. When the total cache capacity drops to $0.1$, \textit{RippleClassic} delivers an average bitrate 9.4\% better than \textit{CE2} with LFU. \textit{RippleFinder} results in a similar performance as \textit{CE2} with LFU across all tested cache capacity, with one exception. At $\omega=0.1$, \textit{RippleFinder} delivers an average bitrate 3.9\% lower than \textit{CE2} with LFU. Later observations show that \textit{RippleFinder} magnifies such small differences by substantially reducing bitrate oscillation irrespective of caching resources.

The trend is similar as popularity skewness, shown in Figure~\ref{Fig:BRZipf}. As expected, average video bitrates increase among all caching schemes 
as the skewness parameter $\alpha$ grows from 0.8 to 1.4, since a greater number of requests target fewer video content.
Here, too, \textit{RippleClassic} and \textit{RippleFinder} will distinguish themselves via improvements in reducing oscillation.

When comparing both \textit{RippleCache}-guided schemes to each other, we observe measurable differences when cache capacity and skew diminish. 
This is explained by the design of \textit{RippleClassic} to optimize the use of available resources against request patterns.  
% when the total amount of cache capacity is constrained, the advantage of optimization stands out which utilizes limited cache resource in a more efficient way than \textit{RippleFinder}. This advantage prevails across different popularity distributions as shown in Figure~\ref{Fig:BRZipf} when we fix the cache capacity at $\omega = 0.2$.  
In contrast the advantages of optimization over popularity-based schemes, including \textit{RippleFinder}, diminish as capacity resources grow or request patterns become predictable. 
% when cache capacity is sufficient, where any popularity-based scheme can already capture request pattern and deliver high-quality video content.

\subsection{Bitrate Switch Count}\label{subsec:impactOscillation}
The \textit{Bitrate Switch Count} measures oscillation by recording the frequency of bitrate switches (including both upgrade and downgrade) in a video session~\cite{dashmetrics}. Results are shown in Figures~\ref{Fig:BOSize} and~\ref{Fig:BOZipf}, as cache capacity and popularity distribution are made to vary, respectively. In all evaluations, both \textit{RippleClassic} and \textit{RippleFinder} reduce bitrate oscillation when compared with popularity-based \textit{CE2} with LFU and \textit{ProbCache}. 
When cache capacity is lowest, or popularity least skewed, \textit{CE2} with LRU appears to meet or exceed \textit{RippleFinder} or \textit{RippleClassic} scheme. The corresponding video bitrate observations for LRU show that this comes at the cost of video quality. The lower video quality for LRU also explains the low degrees of oscillation. Coupled with Figures~\ref{Fig:BRSize} and~\ref{Fig:BRZipf}, we see that even when $\omega = 0.1$ (or $\alpha = 0.8$), our \textit{RippleCache}-guided schemes are seen to reduce oscillation while sustaining the highest levels of video quality.

Observations in Figures~\ref{Fig:BOSize} and~\ref{Fig:BOZipf} also indicate that \textit{RippleFinder} outperforms \textit{RippleClassic}, despite both adhering to \textit{RippleCache} ideals. 
The performance gap may be explained by their difference in caching decision criteria. Recall that \textit{RippleClassic} implements a reward system that approximates adaptation behaviour. This has the effect of optimizing for the consumer's criteria, namely maximal sustainable bitrate. Conversely, the use of utility in \textit{RippleFinder} embeds conventional notions of hit ratio, albeit on a per-path basis. This has the effect of stabilizing video throughput across a single logical cache despite being distributed over multiple volumes. The differences in performance between the two \textit{RippleCache} schemes are reflective of their different emphases on consumer vs. cache performance.
% although \textit{RippleFinder} outperforms \textit{RippleClassic} in terms of \textit{Bitrate Switch Count}, \textit{RippleClassic} still delivers the best possible video quality. As \textit{Bitrate Switch Count} includes both bitrate upgrade and downgrade, the slightly worse bitrate oscillation of \textit{RippleClassic} is partly caused by its higher \textit{Average Video Bitrate} (which means more bitrate upgrades).

\subsection{Rebuffer Percentage}\label{subsec:impactFreeze}
Short-term variations in network and system conditions can adversely affect playback before bitrate adaptations are triggered. One such indication is buffer-induced pausing during playback that manifests on-screen as `freezing'. We measure the impact of `freezing' in terms of \textit{rebuffer percentage}, which is the average time spent in a video freezing state over the active time of a video session~\cite{dashmetrics}. Results are shown in Figures~\ref{Fig:FreezeSize} and~\ref{Fig:FreezeZipf}.

Since video playback freezing relates to the access delay of media segments, caching schemes that achieve high hit ratios must be able to deliver segments before they are needed for playback, otherwise the playback will freeze. This can be seen in Figures Figures~\ref{Fig:FreezeSize} and~\ref{Fig:FreezeZipf}, where both \textit{RippleCache}-guided caching schemes outperform the others. As large amount of video segments with high quality would significantly increase the network delay and choke video traffic, \textit{RippleFinder} and \textit{RippleClassic} reduce system-wide traffic load by satisfying high-bitrate requests as early as possible. Only when the request distribution is least skewed ($\alpha = 0.8$) or there exists limited cache capacity ($\omega = 0.1$), does \emph{RippleFinder} or \textit{RippleClassic} performance diminish to a degree matched by popularity-based schemes. 

Intuitively, \textit{Average Video Bitrate} and \textit{Rebuffer Percentage} are conflicting measures, i.e, a higher video bitrate probably leads to a worse playback freezing. However, simulation results from 
Figures~\ref{Fig:SizeEvals} and~\ref{Fig:ZipfEvals} 
imply that the relationship between these two metrics is more subtle. In support of intuition, for example, at cache capacity $\omega = 0.1$ LFU delivers higher video quality than LRU, but causes almost twice playback freezing. The perceived relationship between metrics is broken when comparing \textit{RippleFinder} at the same $\omega = 0.1$. Here, \textit{RippleFinder} delivers the higher video quality matching LFU but maintains the same rebuffer ratio as LRU. 
%To the end, the chance of freezing is influenced by cache utilization. It is possible for a caching scheme (e.g., \textit{RippleFinder} and \textit{RippleClassic}) to achieve high video quality while minimize the playback freezing. 
%Only when there are insufficient caching resources, the importance of cache utilization diminishes and then the observed results may present a trend that matches the intuition.
Collectively these observations reinforce that, in distributed multimedia caching systems, cache hits have value only if their occurrence is useful to the consumer.

\subsection{Evaluation On A Realistic Topology}
\label{subsec:impactTopology}
\begin{figure*}[!t]
	\centering
	\begin{minipage}{\textwidth}
	    \centering
	    \subfloat[Average Video Bitrate]
	    {\includegraphics[width=0.32\textwidth]{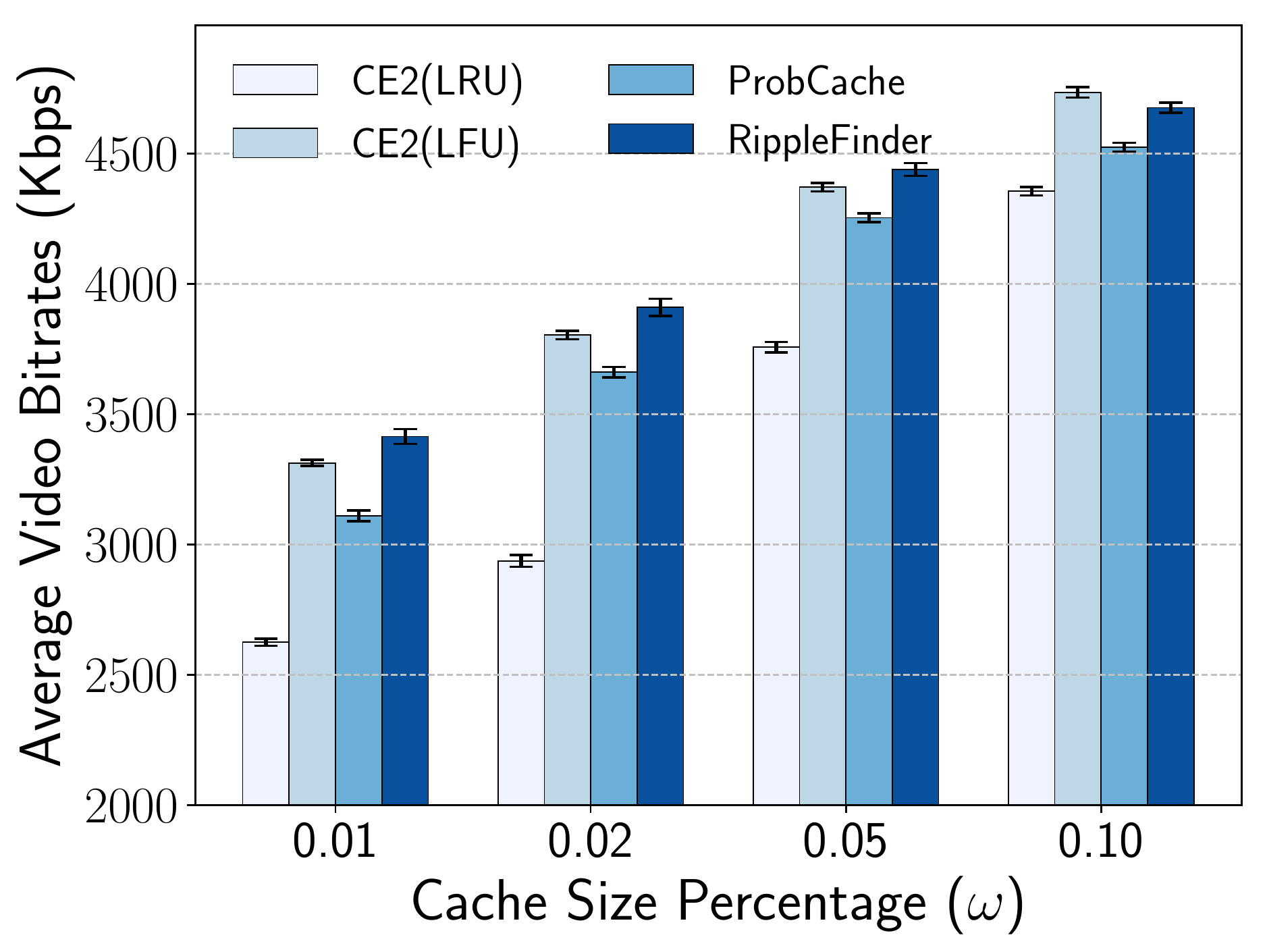}
		\label{Fig:BRSizeTopo}}
		\hfil
		\subfloat[Bitrate Switch Count]
		{\includegraphics[width=0.32\textwidth]{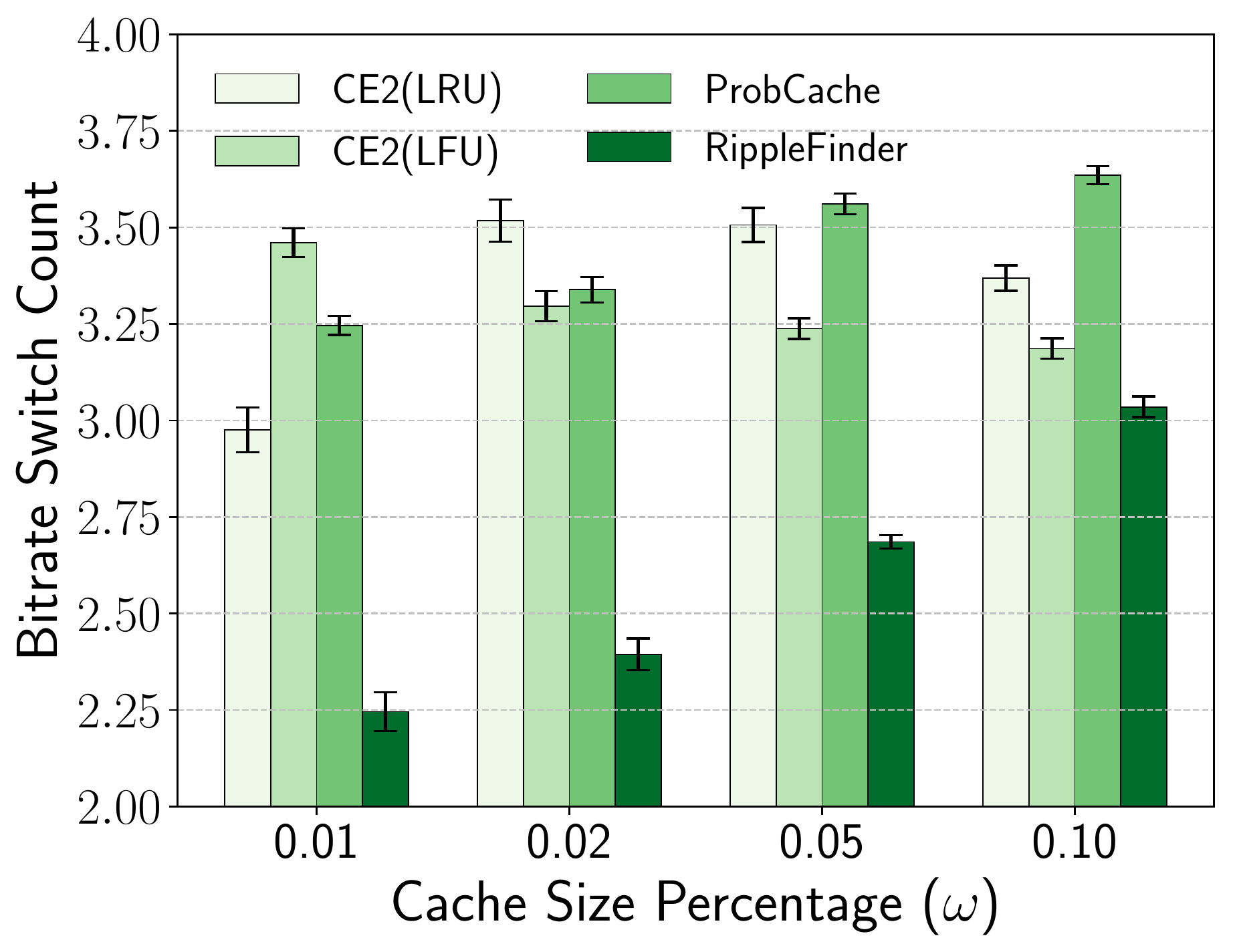}
		\label{Fig:BOSizeTopo}}
		\hfil
		\subfloat[Rebuffer Percentage]
		{\includegraphics[width=0.32\textwidth]{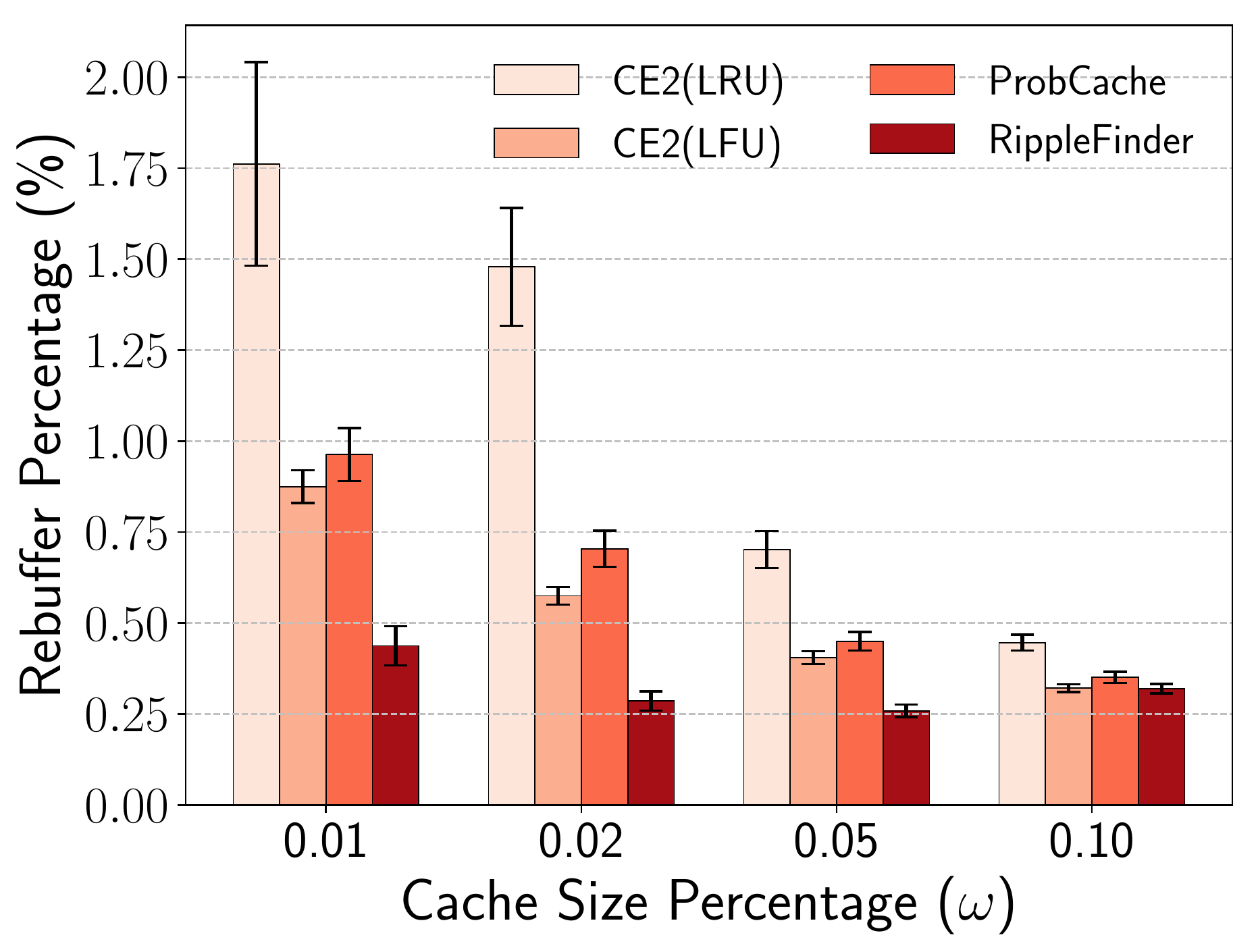}
		\label{Fig:FreezeSizeTopo}}
		\caption{Content Store Size Percentage ($\omega$) for `Large-scale' settings}
		\label{Fig:SizeEvalsTopo}
	\end{minipage}
	\begin{minipage}{\textwidth}
	    \centering
		\subfloat[Average Video Bitrate]
		{\includegraphics[width=0.32\textwidth]{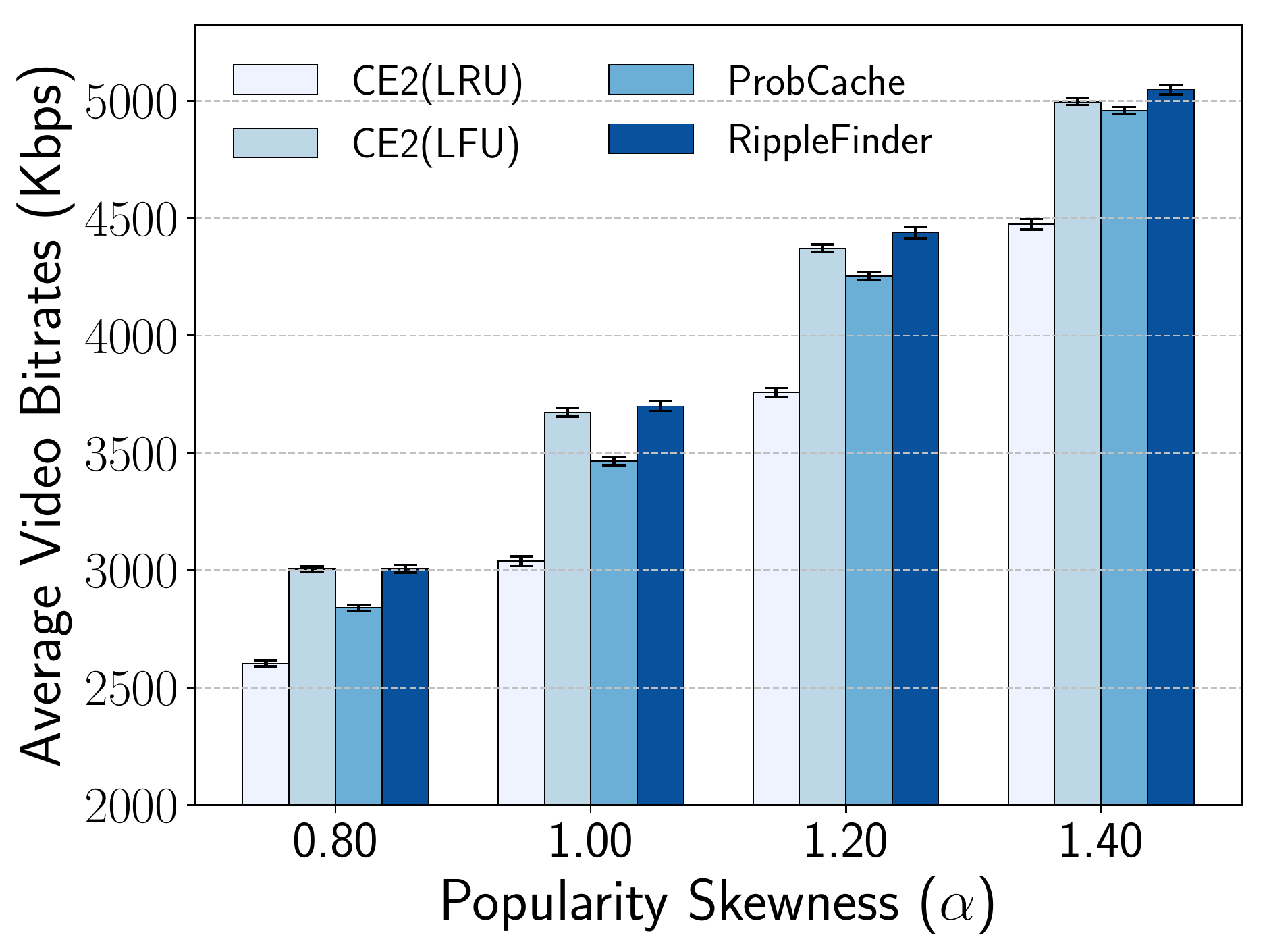}
		\label{Fig:BRZipfTopo}}
		\hfil
		\subfloat[Bitrate Switch Count]
		{\includegraphics[width=0.32\textwidth]{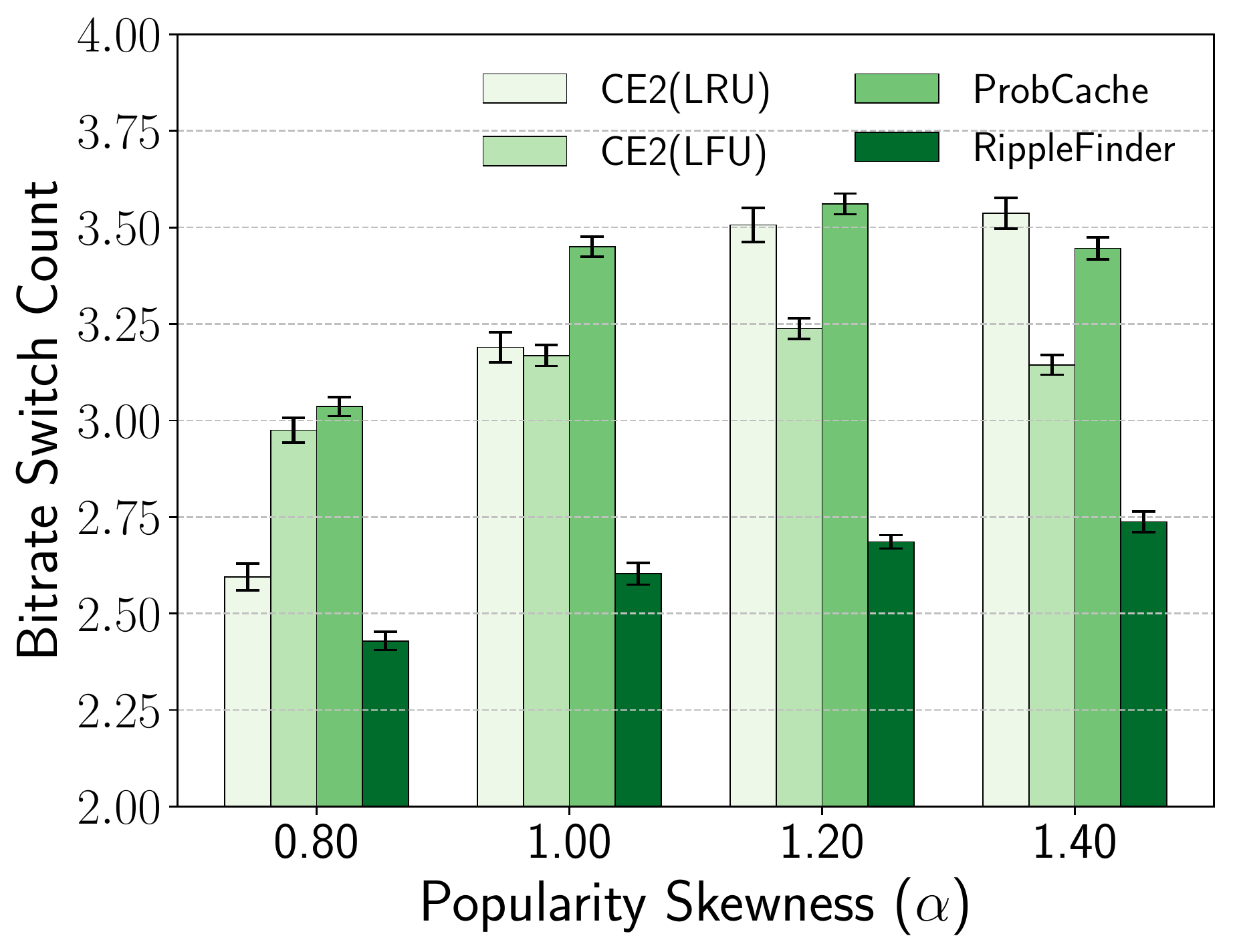}
		\label{Fig:BOZipfTopo}}
		\hfil
		\subfloat[Rebuffer Percentage]
		{\includegraphics[width=0.32\textwidth]{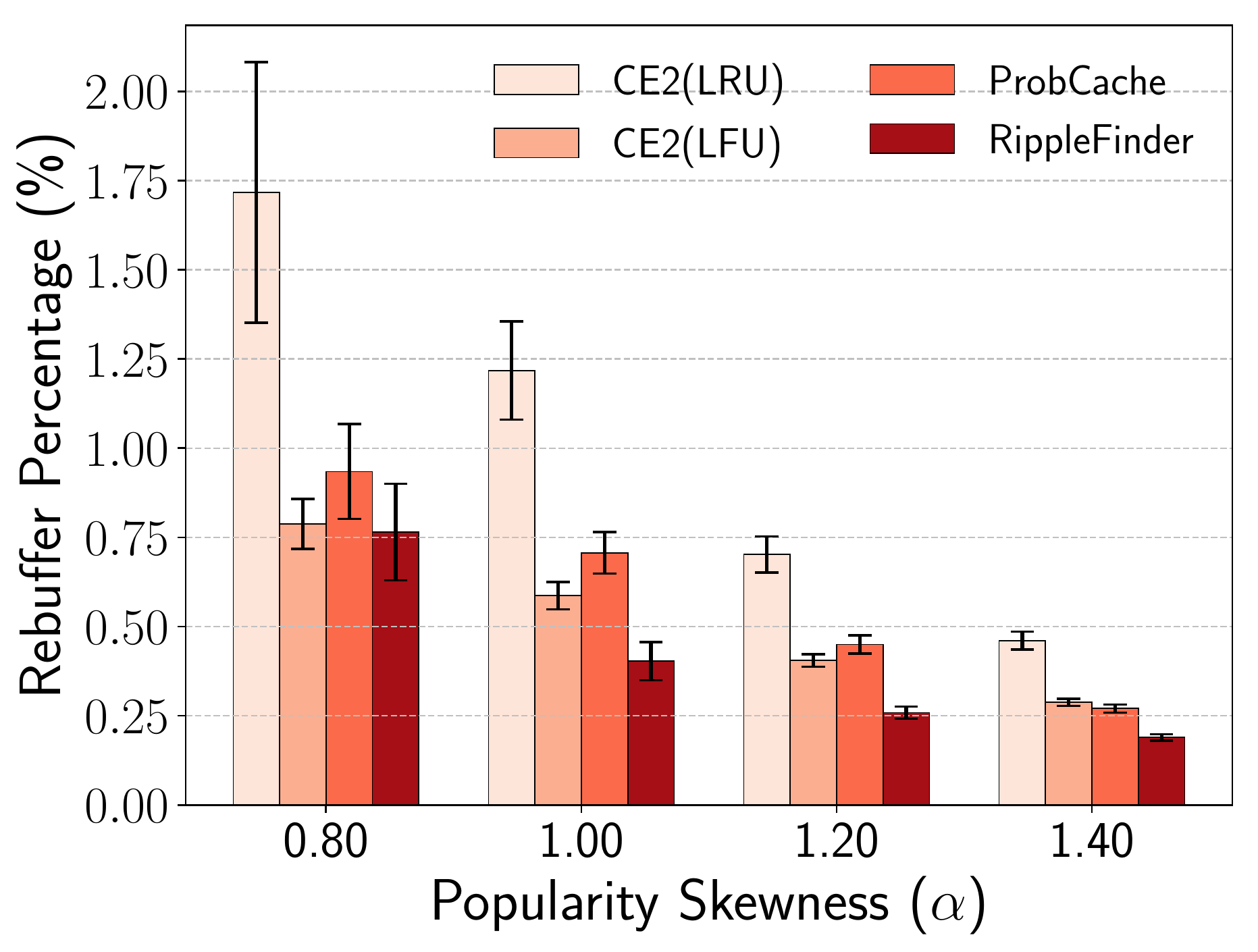}
		\label{Fig:FreezeZipfTopo}}
        \caption{Popularity Skewness ($\alpha$) for `Large-scale' settings}
        \label{Fig:ZipfEvalsTopo}
    \end{minipage}
\end{figure*}
We evaluate over a large 42-node autonomous system (AS) topology generated using BRITE~\cite{medina2001brite}. 
% representative use BRITE~\cite{medina2001brite} to generate a 42-node topology. 
The Barab\'{a}si-Albert (BA) model is first selected to build an autonomous system(AS)-level structure. Connections between ICN routers within each AS are established randomly. A total of 84 video consumers are connected to this network, and request for video content from three producers. Each producer provides 500 video files, each consisting of 50 segments. Remaining simulation settings for this large-scale scenario are listed in Table~\ref{Table:Parameters}.

%shows the potential of \textit{RippleFinder} to handle a scenario with severe conflicting interests.
%no pattern appears for bitrate oscillation: it composes increase and decrease
% Users' QoE under this topology is presented in 
Results in Figures~\ref{Fig:SizeEvalsTopo} and~\ref{Fig:ZipfEvalsTopo} show that \textit{RippleFinder} performance trends are similar to previous observations. In particular, \textit{RippleFinder} meets or exceeds competing levels of video quality and rebuffering, while significantly reducing bitrate oscillation.

This consistent performance of \emph{RippleCache}-guided design across topologies is also noteworthy. When compared to trends of the smaller network captured in Figures~\ref{Fig:SizeEvals} and~\ref{Fig:ZipfEvals}, the performance of competing schemes appears to be affected by size and topology. For example, Figures~\ref{Fig:FreezeSize} and~\ref{Fig:FreezeZipf} show \textit{CE2} with higher rates of rebuffering for LFU than LRU at small $\omega$ or small $\alpha$, respectively. However, in the representative topology LFU and LRU performance is inverted, as can be seen in Figures~~\ref{Fig:FreezeSizeTopo} and~\ref{Fig:FreezeZipfTopo}. 
% This is because the playback freezing (and also oscillation) is a complex consequence that is made by the interplay between caching and adaptation control. 
These differences further demonstrate the poorly understood interactions between caching and adaptation controls.
% In this simulation settings, a large-scale  network topology and multiple video producers would result in a more realistic traffic pattern that triggers a different reaction by caching schemes. 
The consistent QoE performance delivered by \textit{RippleFinder} across different traffic patterns and topologies is important for real-world deployments.
% Our proposed caching scheme, \textit{RippleFinder}, presents a consistent improvement on QoE 

\subsection{Discussion of Results}\label{SubSec:Discussion}
Throughout our evaluations we notice the ability of \textit{CE2} with LFU in terms of delivered video quality and playback freezing. Looking ahead, the robustness of LFU suggests that performance gains promised by ICNs specifically, and caching hierarchies generally, may be dependent on on their ability to exploit content characteristics. Otherwise caching mechanisms may be mooted by simple popularity, alone, and the corresponding simplicity of LFU.

The general hypothesis that cache placement should be informed by content characteristics is reinforced by \emph{RippleFinder}/\textit{RippleClassic} observations. By designing a cache placement scheme for adaptive streaming content, we draw insights that run counter to convention. Lower quality content that is pushed into the core, for example, can improve end-user QoE. Edge caches are left with additional capacity for higher-quality content. Consequently, content quality at all bitrates becomes network- rather than cache-limited.
\section{Conclusion}\label{Sec:Conclusion}
In this paper, we have argued that ICN cache placement should be tailored for adaptive streaming, as bitrate adaptation mechanisms appear to clash with generic ICN caching techniques.  We highlight the issue of oscillation dynamics which is caused by the interplay between in-network caching and bitrate adaptation control, and present a primer in a novel approach to caching, and establishes the premise of safe guarding cache partitions for higher bit-rates, allowing for more ideal cache placement strategies for adaptive video content.

Our proposed safe-guarding mechanism enforces bitrate-based partitioning of cache capacities, named as \textit{RippleCache}, in order to stabilize bandwidth fluctuation. In \textit{RippleCache}, a network of caches is viewed along each forwarding path from consumers, where the essence is safeguarding high-bitrate content on the edge and pushing low-bitrate content into the network core. To validate the concept and demonstrate the potential gain of \textit{RippleCache}, we implement two cache placement schemes, \textit{RippleClassic} and \textit{RippleFinder}, where our experiment results contrast to leading caching schemes, and demonstrate how cache partitioning would improve users' QoE, in terms of high video quality and significant reduction on bitrate oscillation.

More importantly, our explorations yield the following conclusions:
\begin{inparaenum}
	\item The operational mandate of bitrate adaptation algorithms significantly impacts in-network caching schemes, thus caching must seamlessly cooperate with adaptation. Existing schemes that apply a snapshot approach cannot be applied directly for adaptive streaming application, as they ignore the need of cooperation, which results in severe bitrate oscillation.
	\item The problem of bitrate oscillation can be tackled by concatenating caches along a forwarding path into a \textit{cache path}. Although cache hits on a standalone router would result in similar throughput, adaptation-level dynamics vary across encoding bitrates such that even exact same throughput will not bring the same adaptation decision for each bitrate. By zooming out our view from one cache to the range of a forwarding path, we can arrange video content of different bitrates at a hop distance from consumers that maximizes the chance of maintaining the same adaptation decision, which is the key to avoid bitrate oscillation.
	\item It is possible for a caching scheme to deliver video consumers high-quality content while ensuring near-zero playback freezing and minimal bitrate oscillation. Our experiments demonstrate that there is significant room of improvement for future caching policies to enhance QoE by practicing cache partitioning and inheriting from \textit{RippleCache} principle.
\end{inparaenum}
This study paves the way for caching schemes that can interact with bitrate selection algorithms, and handle the dependency between adaptation control and caching via network prediction for future request patterns.

\bibliographystyle{IEEEtran}
\bibliography{RippleCache}

\end{document}